\documentstyle[12pt]{article}
\newcommand{\sect}[1]{\setcounter{equation}{0}\section{#1}}

\def\rf#1{(\ref{eq:#1})}
\def\lab#1{\label{eq:#1}}
\def\nn{\nonumber \\}
\newcommand{\beano}{\begin{eqnarray*}}
\newcommand{\enano}{\end{eqnarray*}}
\def\bea{\begin{eqnarray}}
\def\ena{\end{eqnarray}}
\def\be{\begin{equation}}
\def\ee{\end{equation}}

\relax
\newcommand{\bz}{{\bar{0}}}
\newcommand{\bo}{{\bar{1}}}
\newcommand{\hf}{\frac{1}{2}}
\font\fld=msbm10 at 12 pt

\newcommand{\fl}[1]{\mbox{\fld #1}}     
\newcommand\partder[2]{{{\partial {#1}}\over{\partial {#2}}}}

\newcommand{\sm}{\mbox{-}}  
\def\a{\alpha}
\def\b{\beta}
\def\d{\delta}

\def\ve{\varepsilon}
\def\g{\gamma}

\def\l{\lambda}
\def\L{\Lambda}

\def\pa{\partial}

\def\t{\tau}
\def\th{\theta}

\def\ca{{\cal A}}

\def\cg{{\cal G}}
\def\ch{{\cal H}}
\def\ck{{\cal K}}
\def\cl{{\cal L}}

\def\cz{{\cal Z}}

\newcommand{\ad}{\mbox{\rm ad}}

\newcommand{\str}{\mbox{\rm str}\hspace{2pt}}
\newcommand{\tr}{\mbox{\rm tr}\hspace{2pt}}
\newcommand{\one}{1 \hspace{-3pt}{\rm I}}
\newtheorem{defi}{Definition}
\newtheorem{theo}{Theorem}

\newtheorem{lem}{Lemma}
\newtheorem{cor}{Corollary}
\def\NPB#1#2#3{{\sl Nucl. Phys.} {\bf B#1} (#2) #3}

\def\CMP#1#2#3{{\sl Commun. Math. Phys.} {\bf #1} (#2) #3}

\def\PLA#1#2#3{{\sl Phys. Lett.} {\bf #1A} (#2) #3}
\def\PLB#1#2#3{{\sl Phys. Lett.} {\bf #1B} (#2) #3}
\def\JMP#1#2#3{{\sl J. Math. Phys.} {\bf #1} (#2) #3}

\def\RMaP#1#2#3{{\sl Reports on Math. Phys.} {\bf #1} (#2) #3}

\def\JPA#1#2#3{{\sl J. Physics} {\bf A#1} (#2) #3}
\def\JSM#1#2#3{{\sl J. Soviet Math.} {\bf #1} (#2) #3}
\def\MPLA#1#2#3{{\sl Mod. Phys. Lett.} {\bf A#1} (#2) #3}

\newtheorem{Prop}{Proposition}
\begin{document}
%
\begin{titlepage}
\vspace{-1cm}
\noindent
\hfill{US-FT/10-99}\\
\vspace{0.0cm}
\hfill{hep-th/9905103}\\
\vspace{0.0cm}
\hfill{May 1999}\\
\phantom{bla}
\vfill
\begin{center}
{\large\bf  Non-local conservation laws and flow  equations}\\
\vspace{0.3cm}
{\large\bf for supersymmetric integrable hierarchies}
\end{center}

\vspace{0.3cm}
\begin{center}
Jens Ole Madsen~ and~ J. Luis Miramontes
\par \vskip .1in \noindent
{\em 
Departamento de F\'\i sica de Part\'\i culas,\\
Facultad de F\'\i sica\\
Universidad de Santiago de Compostela\\
E-15706 Santiago de Compostela, Spain}\\
\par \vskip .1in \noindent
madsen@fpaxp1.usc.es~~~~ miramont@fpaxp1.usc.es
\normalsize
\end{center}
\vspace{.2in}
\begin{abstract}
\vspace{.3 cm}
\small
\par \vskip .1in \noindent
An infinite series of Grassmann-odd and Grassmann-even flow equations
is defined for a class of supersymmetric integrable hierarchies
associated with loop superalgebras. All these flows commute with the
mutually commuting bosonic ones originally considered to define these
hierarchies and, hence, provide extra fermionic and bosonic symmetries
that include the built-in $N=1$ supersymmetry transformation. The
corresponding non-local conserved quantities are also constructed. As an
example, the particular case of the principal supersymmetric hierarchies
associated with the affine superalgebras with a fermionic simple root
system is discussed in detail.

\end{abstract}
\vfill
\end{titlepage}
\sect{Introduction}

~\indent
Arguably, one of the most important works in the subject of
integrable systems was that of  Drinfel'd and Sokolov~\cite{DrSo}, who
showed how to associate integrable hierarchies of
zero-curvature equations with the loop algebra of an affine Lie
algebra. Their construction and its
generalizations~\cite{miramontesetal} provide a systematic
approach to the study and classification of many integrable
hierarchies previously described by means of pseudo-differential
Lax operators~\cite{Pseudo}.

It is not difficult to extend the generalized
Drinfel'd-Sokolov construction to the case of superalgebras. One
simply has to replace the loop algebra by a loop
superalgebra and include fermionic (Grassmann-odd) anticommuting
fields among the dynamical degrees of freedom. However, the resulting
hierarchy will not necessarily be supersymmetric. The first authors
who succeeded in finding a supersymmetric generalization of the DS
construction were Inami and Kanno~\cite{InKa1} who restricted to the
class of affine superalgebras with a fermionic simple root system.
Since they made use of the principal gradation of the loop
superalgebra in an essential way, their work has to be viewed as the
direct generalization of the original DS construction. More recently,
Delduc and Gallot~\cite{DelGal} realized that it is possible to
associate a supersymmetric integrable hierarchy of the DS type with 
each constant graded odd element $\Psi$ of the loop superalgebra whose
square $\Lambda = [\Psi, \Psi]/2$ is semi-simple, a condition that
is obviously satisfied in the cases considered by Inami and Kanno. 

A common feature of the hierarchies constructed in~\cite{InKa1}
and~\cite{DelGal} is that they consist only of bosonic flow
equations. This is in sharp contrast with the supersymmetric extensions
of the KP hierarchy (SKP)  which always include both Grassmann-odd
and Grassmann-even flow equations~\cite{ManRad}. This observation led
Kersten to find and infinite set of fermionic non-local conservation
laws~\cite{Kersten} for the supersymmetric extension of the KdV
equation obtained by Manin and Radul as a reduction of the
SKP hierarchy~\cite{ManRad}. A few years later, Dargis
and Mathieu showed that they actually generate an infinite sequence of
non-local Grassmann-odd flows~\cite{DaMa} (see also~\cite{Ra}).

The aim of this paper is to show that the same is true for the whole
class of supersymmetric hierarchies of~\cite{DelGal} by
constructing an infinite series of non-local (bosonic and fermionic)
flow equations and conserved quantities which generalize those obtained
in~\cite{Kersten,DaMa} for the supersymmetric KdV equation. The new
flows close a non-abelian superalgebra that has to be regarded as an
algebra of symmetry transformations for the hierarchy. It includes
the built-in $N=1$ supersymmetry transformation and, in some cases,
extended supersymmetry transformations. 

Following ref.~\cite{DelGal}, a generalized supersymmetric hierarchy of
equations can be associated with a fermionic Lax operator of the form
${\cal L} = D +q(x,\theta) + \Psi$, where $q(x,\theta)$ is an $N=1$
Grassmann-odd superfield taking values in a particular subspace of the
loop superalgebra, $D$ is the superderivative, and $\Psi$ is a constant
graded odd element whose square $\Lambda = [\Psi, \Psi]/2$ is
semi-simple. Delduc and Gallot defined an infinite set of mutually
commuting bosonic flows and conserved quantities associated with the
elements in the centre of ${\cal K}= {\rm Ker\/}({\rm ad\>}\Lambda)$,
which contains only even elements. In contrast, in our construction,
there will be a non-local flow equation and conserved quantity for
each (fermionic or bosonic) element in $\cal K$ with non-negative
grade. These flows close a non-abelian superalgebra isomorphic to
the subalgebra of $\cal K$ formed by the elements with non-negative
grade.

The authors of~\cite{miramontesetal} distinguished between
(bosonic) generalized DS hierarchies of type-I and type-II. The
generalized DS hierarchies are associated with bosonic Lax operators
of the form $L= \partial_x + Q(x) -\Lambda$, where $Q(x)$ is a bosonic
field taking values in a subspace of a loop algebra and $\Lambda$ is a
constant semi-simple graded element. Then, the hierarchy is of type-I
or type-II depending on whether $\Lambda$ is regular or not, {\em
i.e.\/}, a type-II hierarchy is associated with a semi-simple element
$\Lambda$ such that ${\cal K}= {\rm Ker\/}({\rm ad\>}\Lambda)$ is
non-abelian. In~\cite{DelGal}, the analogue of $L$ is the even Lax
operator 
$[\hat{\cal L} , {\cal L}]/2 = D^2 + Q(x,\theta) - \Lambda $.
Therefore, from this point of view, all the supersymmetric hierarchies
of~\cite{DelGal} have to be considered as type-II: $\Psi$ is in
$\cal K$ but $[\Psi, \Psi]= 2\Lambda \not=0$, which proves that 
$\cal K$ is non-abelian. In fact, it is straightforward to extend our
construction of non-local flow equations and conserved quantities to
all the type-II bosonic hierarchies of~\cite{miramontesetal}.

The paper is organized as follows. In section~\ref{sec2} we
briefly summarize the supersymmetric hierarchies of Delduc and Gallot,
making particular emphasis on the possibility of having fermionic local
conserved quantities. In the next two sections we present the
construction of non-local flows, sec.~\ref{sec3}, and conserved
quantities, sec.~\ref{sec4}. We also characterise the flows which are
compatible with the supersymmetry transformation, and the conserved
quantities which are supersymmetric. As an example, in
section~\ref{sec-5} we give a very detailed description of the
non-local flows and conserved quantities for the principal
hierarchies originally considered by Inami and Kanno, {\em
i.e.\/}, those associated with superalgebras with a fermionic
simple root system. We show that the complete set of
local and non-local flow equations are associated with a
subalgebra of the superoscillator algebra constructed by Kac
and van de Leur in~\cite{KacLeur}, which is the principal `super
Heisenberg algebra' of $a_{\infty|\infty}$~\cite{Fock}. This relationship
generalizes the well known role of the principal Heisenberg algebra in
the original DS hierarchies, which could be used to derive a
$\tau$-function formalism for these hierarchies following the method
of~\cite{Tau,Wilson,KW} and, especially, to construct solutions for the
equations of the hierarchy using super vertex operator
representations. It is also remarkable that, in the particular case of
the affine superalgebras $A(m,m)^{(1)}$, there is a local
Grassmann-odd flow $\bar{D}_{1}$ that, together with
$D$, closes an $N=2$ supersymmetry algebra, in agreement with the results
of~\cite{InKa2,DelGal}. In section~\ref{SKdV} we apply our construction
to the supersymmetric KdV equation in order to show how the results
of~\cite{DaMa} are recovered. As a bonus, we get additional non-local
conserved quantities not obtained in previous works. Our conventions and
some basic properties of Lie superalgebras are presented in appendix~B.
Since we are not aware of any reference where they are available, we
have included in this appendix detailed expressions for the 
matrix-representations of the affine Lie superalgebras with fermionic
simple root systems which are needed for the understanding of
section~\ref{sec-5}. Recall that this class of superalgebras play an
important role in supersymmetric Toda theories and supersymmetric
hamiltonian reduction and, hence, these expressions should be useful
beyond the scope of this paper. Our conclusions are presented in
section~\ref{sec7}.

\sect{Review of the Delduc-Gallot construction}
\label{sec2}

~\indent
Following~\cite{DelGal}, a supersymmetric (partially modified) KdV
system  can be associated with four data $({\cal A},
d_1,d_0,\Psi)$. The first, $\cal A$, is a twisted loop superalgebra
\be
{\cal A} = {\cal L}({\cal G}, \tau) \subset {\cal G}\otimes
\fl{C}[\lambda, \lambda^{-1}]
\lab{algebra}
\ee
attached to a finite dimensional classical Lie superalgebra
$\cal G$ together with an automorphism $\tau$ of finite order. The
second and third, $d_1$ and $d_0$, are derivations of $\cal A$
that induce two compatible integer gradations: $[d_0, d_1]=0$ .
Then, the subsets ${\cal A}^{n}_m = {\cal A}_{m}
\cap {\cal A}^{n} $, where ${\cal A}_m = 
\{X\in {\cal A} \mid [d_0 , X] = m X\} $ and ${\cal A}^n = 
\{X\in {\cal A} \mid [d_1 , X] = n X\} $, define a bi-grading of
$\cal A$. We will assume that
\be
{\cal A}^{<0} \subset {\cal A}_{\leq 0}, \quad
{\cal A}^{0} \subset {\cal A}_{0}, \quad
{\cal A}^{>0} \subset {\cal A}_{\geq 0}\>,
\lab{Include}
\ee
which means, in the notation of~\cite{DelGal}, that we restrict
ourselves to the `KdV type systems'. 

Finally, $\Psi$ is a constant odd element of $\cal A$ with
positive $d_1$-grade, {\it i.e.\/}, $[d_1, \Psi] = k\Psi$ with
$k>0$, whose square $\Lambda ={1\over2}[\Psi, \Psi]$ is
semi-simple:
\be 
{\cal A} = {\rm Ker\/}({\rm ad\>}\Lambda) \oplus {\rm Im\/}({\rm
ad\>}\Lambda) \>.
\ee
Moreover, $\Psi$ has to satisfy the non-degeneracy
condition~\footnote{Notice that our non-degeneracy condition
implies that ${\rm Ker\/}({\rm ad\>}\Lambda) \cap {\cal A}_{0}^{<0}
= {0}$, which is the non-degeneracy condition mentioned
in~\cite{DelGal}. However, the latter is unnecessarily more
restrictive that the former.} 
\be
{\rm Ker\/}({\rm ad\>}\Psi) \cap {\cal A}_{0}^{<0} = {0}\>,
\lab{Degen}
\ee
and it is worth recalling that 
\be
[{\rm Ker\/}({\rm ad\>}\Lambda) , {\rm Ker\/}({\rm ad\>}\Lambda)]
\subset {\rm Ker\/}({\rm ad\>}\Lambda) \>,
\qquad [{\rm Ker\/}({\rm ad\>}\Lambda), {\rm Im\/}({\rm
ad\>}\Lambda)] \subset {\rm Im\/}({\rm ad\>}\Lambda)\>.
\ee
Therefore, ${\rm Ker\/}({\rm ad\>}\Lambda)$ is a, 
generally non-abelian, subalgebra of $\cal A$ and we will denote by
$\cal Z$ its centre, which  contains only even elements of $\cal
A$. 
 
The dynamical degrees of freedom will be superfields taking values in
the tensor product
$\underline{\cal A}$ of $\cal A$ with some Grassmann algebra
${\cal G}r = {\cal G}r_{\overline 0} \oplus {\cal G}r_{\overline
1}$. An element in this space is called {\em even} if it belongs to
$_+{\cal A} = {\cal A}_{\overline 0}\otimes{\cal G}r_{\overline
0}\> \oplus\> {\cal A}_{\overline 1}\otimes{\cal G}r_{\overline
1}$, and {\em odd} if it belongs to $_-{\cal A} = {\cal
A}_{\overline 0}\otimes{\cal G}r_{\overline 1}\> \oplus\> {\cal
A}_{\overline 1}\otimes{\cal G}r_{\overline 0}$. Let $T^a$ be a
set of basis vectors of $\cal A$. Then, the (super) commutator of two
elements $A = \sum_a A_a T^a$ and $B= \sum_a B_a T^a$ of
$\underline{\cal A}$ will be defined as
\be
[A, B] = \sum_{a,b} A_a B_ b [T^a, T^b]\>,
\lab{SC}
\ee
where $A_a, B_b \in {\cal G}r$. With this definition, the
commutator satisfies the symmetry properties listed in
table~\ref{SYM}, where we denote by $\hat{\;}$ (hat) the automorphism of
$\cal A$ that changes the sign of odd elements:
\be
M= M_{\overline0} + M_{\overline1}\in {\cal A}\>, \qquad
\hat{M} = M_{\overline0} - M_{\overline1}\>.
\ee

\begin{table}
\begin{center}
\begin{tabular}{|r|r|r|}
\hline 
$[A\>,\> B]\> =\>$ & $A\in {_+{\cal A}}$ & $A\in {_-{\cal A}}$ \\
\hline
$B\in {_+{\cal A}}$ & $-\> [B\>,\> A]$ & $-\> [\hat B\>,\> A]$ \\
\hline
$B\in {_-{\cal A}}$ & $-\> [B\>,\> \hat A]$ & $+\> [\hat B\>,\> \hat
A]$
\\ 
\hline 
\end{tabular}
\end{center}
\caption{Symmetry properties of the supercommutator.}
\label{SYM}
\end{table}

Consider an $N=1$ superspace
with coordinates $\tilde x=(x, \theta)$ with the supersymmetric
covariant derivative defined in the usual way:
\be
D= {\partial\over \partial \theta} + \theta {\partial\over
\partial x}\>, \qquad D^2 = {\partial\over \partial x}\>.
\lab{SupD}
\ee
Then, let us introduce the odd Lax operator
\be
{\cal L} = D + q + \Psi\>,
\lab{OddLax}
\ee
where $q = q(\tilde x)$ is an odd superfield that takes values in
$_-{\cal A}_{\geq0}^{<k}$. From the odd Lax
operator $\cal L$ we can obtain the following even operator
\be
L ={1\over2} [\hat{\cal L} , {\cal L}] = \partial + Q
- \Lambda\>,
\lab{EvenLax}
\ee
where the even superfield
\be
Q = Q(\tilde x) = Dq + {1\over2}[\hat q , q] - [\Psi, q]
\lab{EvenQ}
\ee
takes values in $_+{\cal A}_{\geq0}^{<2k}$.

The phase space of the hierarchy will be ${\cal
Q}/{\cal N}$, where 
\be
{\cal Q} = \{q(\tilde x) \mid q(\tilde x) \in {_-{\cal
A}}_{\geq0}^{<k}\}
\lab{Space}
\ee
and $\cal N$ is the group of gauge
transformations ${\rm e\/}^\gamma$ acting on $\cal Q$ according 
to
\be
{\rm e\/}^\gamma : {\cal L} \longmapsto {\rm e\/}^{\hat \gamma}
{\cal L} {\rm e\/}^{-\gamma}
\lab{Gauge}
\ee
for any superfield $\gamma= \gamma(\tilde x)$ taking
values in $_+{\cal A}_{0}^{<0}$. Let us define a
vector subspace $V$ of $\cal A$ such that
\be
{\cal A}_{\geq0}^{<k} = [\Psi, {\cal A}^{<0}_0]
\oplus V\>.
\ee
Then, due to the non-degeneracy condition~\rf{Degen}, for any
$q\in {\cal Q}$ there is a unique $q$-dependent gauge
transformation such that the image $q_V= q_V(\tilde x)$ is in
\be
{\cal Q}_V = \{q(\tilde x) \mid q(\tilde x) \in {_-V}\}\>;
\lab{SpaceDS}
\ee
in other words, ${\cal Q}_V$ is isomorphic to ${\cal Q}/{\cal N}$.
When regarded as functions of $q\in \cal Q$, the components of
$q_V$ are super-differential polynomials that freely generate the set
of gauge invariant polynomials on $\cal Q$, and ${\cal Q}_V$ is
said to specify a Drinfel'd-Sokolov gauge.

The construction of~\cite{DelGal} relies on the following
version of the dressing procedure: 
\begin{lem}
\label{lem1}
For any $q\in {\cal Q}$, there exists a unique $F\in {_+({\rm
Im\/}({\rm ad\/}\Lambda))}^{<0}$ such that
\be
{\cal L}_0 = {\rm e\/}^{\hat F} {\cal L} {\rm e\/}^{- F} =
D + \Psi + H\>,
\lab{Abelianize}
\ee
where $H\in {_-({\rm Ker\/}({\rm ad\>}\Lambda))}^{<k}$. Moreover,
the components of $F$ and $H$ are super-differential polynomials
of~$q$.
\end{lem}

\noindent
This result is used to define an infinite number of
commuting flows on ${\cal Q}/{\cal N}$ associated with the
elements of ${\cal Z}^{\geq0}$ as follows. For each constant $b \in
{\cal Z}^{\geq0}$, let us define the even superfield $B_b = {\rm
e\/}^{-\hat F}  b {\rm e\/}^{\hat F}$. Then, the flow on the space
of gauge invariant functionals of $q$ is induced
by~\footnote{Since the flows are defined
on the set of gauge invariant functions of $q(\tilde x)$, there
are many equivalent definitions of $\partial_b q$ corresponding to
the same flow. Namely, one can take $\partial_b q = [{\cal R}(B_b) +
\gamma, {\cal L}]$,  where $\gamma$ is an arbitrary superfield in ${_+{\cal
A}}_{0}^{<0}$ that generates an infinitesimal gauge
transformation.}
\bea
{\partial \over \partial t_b} q = \partial_b q&= & [{\cal
P}_{\geq0}(B_b), {\cal L}]  = - [{\cal P}_{<0}(B_b), {\cal L}]
\nn  &=& [{\cal R}(B_b), {\cal L}]\>,  
\lab{LocFlow}
\ena
where ${\cal P}_{\geq0}$ and ${\cal P}_{<0}$ are the projectors on
the subalgebras
${\cal A}_{\geq0}$ and ${\cal A}_{<0}$, respectively, and 
${\cal R} = {1\over2}({\cal P}_{\geq0} - {\cal P}_{<0})$ is a
(super) classical r-matrix.

In the following, it will be useful to introduce the notation ${\cal
K} = {\rm Ker\/}({\rm ad\>}\Lambda)$. Let us consider the evolution
of $H$ in~\rf{Abelianize}, which reads
\be
\partial_b H = - D{\hat A_b} + [A_b , \Psi
+H]\>,
\ee
where $A_b$ is a super-differential polynomial of $q$ taking values
in $_-{\cal K}^{<0}$. Following the work of Inami and
Kanno~\cite{InKa1}, let
$\widetilde{\cal K}$ be a vector subspace of $\cal K$ such that
\be
{\cal K} = [{\cal K} , {\cal K}] \oplus \widetilde{\cal K}\>.
\lab{Ktilde}
\ee
Then, an infinite set of local conservation laws can be
associated with the components of $H$ in $\widetilde{\cal K}$.
To be precise, let $[{\cal K}, {\cal K}]^{\bot} =\{b \in {\cal
K} \mid \langle[{\cal K}, {\cal K}], b\rangle=0\}$. Since the
restriction of  the bilinear form to $\cal K$ is non-degenerate,
$[{\cal K}, {\cal K}]^{\bot}$ is conjugate to $\widetilde{\cal K}$
with respect to $\langle\cdot, \cdot\rangle$. Then, for each $\xi \in
[{\cal K}, {\cal K}]^{\bot}$ there is a local conservation law
\be
\pa_b \langle\xi, H\rangle = -D\langle \xi, \hat A_b\rangle.
\lab{ConsL}
\ee 

A particular subset of $[{\cal K}, {\cal K}]^{\bot}$ is $\cal Z$,
the centre of $\cal K$, which obviously satisfies $\langle [{\cal
K}, {\cal K}], {\cal Z}\rangle =0$ due to the invariance of the
bilinear form. Therefore, since
$\cal Z$ contains only even elements of $\cal A$, the components of
$H$ corresponding to ${\cal Z}$ provide an infinite set of
`even' local conserved quantities, which are the Hamiltonians
corresponding to the flows defined by~\rf{LocFlow}. Namely, for
$b\in{\cal Z}^{\geq0}$,  
\be
{\cal H}_b(q) = \int d\tilde x \langle b, H\rangle
\lab{Ham}
\ee
is the Hamiltonian that generates the flow $\partial/\partial
t_b$~\cite{DelGal}.

However, in general, it is not true that ${\cal Z}=
[{\cal K}, {\cal K}]^{\bot}$ or, in other words, that $\cal Z$
and $\widetilde{\cal K}$ are conjugate with respect to
$\langle\cdot, \cdot\rangle$. Consequently, eq.~\rf{ConsL} might
provide more local conservation laws than those associated with
$\cal Z$ and given by~\rf{Ham}. In particular, and in contrast with
$\cal Z$, the vector subspace $[{\cal K}, {\cal K}]^{\bot}$ can
contain odd elements of $\cal A$ and, then, eq.~\rf{ConsL} would
provide Grassmann-odd local conserved quantities. 

All this suggest that it should be possible to generalize the
construction of~\cite{DelGal} to define new flows associated somehow
with the elements of $[{\cal K}, {\cal K}]^{\bot}$ that are not in ${\cal
Z}$. In particular, it should be possible to define fermionic flows
associated with the odd elements of $[{\cal K}, {\cal
K}]^{\bot}$.  

\sect{Non-local flows}
\label{sec3}

~\indent
In this Section, we will associate flow equations with
all the elements of ${\cal K}= {\rm Ker\/}({\rm ad\/} \Lambda)$.
In general, these equations will be non local, {\it i.e.\/}, the
evolution of $q$ will not be given by a super-differential
polynomial of $q$. Moreover, instead of commuting among themselves,
the resulting flows will close a non-abelian superalgebra isomorphic to
${\cal K}^{\geq0}$.

\subsection{Formal dressing transformations}

~\indent
Our construction makes use of a different version of the
formal dressing method inspired by the work of Wilson on the
tau-function approach to hierarchies of KdV type~\cite{Wilson,Tau}.
First of all, for any superfield $\omega\in {_+{\cal K}}^{<0}$,
notice that the transformation 
\be
{\rm e\/}^{F} \longrightarrow {\rm e\/}^{{F'}} = {\rm
e\/}^{\omega} {\rm e\/}^{F}
\ee 
in eq.~\rf{Abelianize} is equivalent to
\be 
{\cal L}_0 \longrightarrow {\rm e\/}^{\hat \omega} {\cal L}_0
{\rm e\/}^{- \omega}=  D + \Psi + H'\>,
\ee
which does not change the form of the right-hand side
of~\rf{Abelianize}. 

\begin{Prop}
\label{Prop1}
The Lax operator $\cal L$ can be `constrained' to ensure
the existence of an even superfield $\omega\in {_+{\cal K}}^{<0}$
such that
\be
{\rm e\/}^{\hat \omega} (D + \Psi + H) {\rm e\/}^{- \omega} = D
+\Psi\>.
\lab{AbelPlus}
\ee
The superfield $\omega$ is a non-local functional of $q$, and the
constraints involve the components of $H$ in ${\cal K}^{\geq-k}$.
\end{Prop}

\noindent
Since the proof is quite involved, it will be presented in Appendix~A.

\vskip .6truecm
It is not easy to give the precise form of the constraints on
$H^{\geq -k}$ required by Proposition~1. However, their
nature can be clarified by considering the
transformation~\rf{AbelPlus} acting on the even Lax operator
\be
L_0 ={1\over2} [\hat{\cal L}_0 , {\cal L}_0] = \partial + h
- \Lambda\>,\qquad
h = DH + {1\over2}[\hat H , H] - [\Psi, H]\>.
\lab{EvenH}
\ee
Then, eq.~\rf{AbelPlus} implies 
\be
{\rm e\/}^{\omega} (\partial + h - \Lambda) {\rm e\/}^{- \omega}
= \partial - \Lambda\>
\ee
that manifests the constraints $h^{\geq0}=0$, which are well
known in the context of the tau-function approach to (bosonic) 
hierarchies of KdV type~\cite{Tau}. 

The relation between the latter constraints and those required by
Proposition~1 can be explicitly shown for a hierarchy
where the $d_1$-grade of $\Psi$ is $k=1$. Then, according
to~\rf{EvenH},
\bea
h^1 &=& -[\Psi, H^0] \nn
h^0 &=& DH^0 + {1\over2}[\hat H^0 , H^0] - [\Psi, H^{-1}] = -[\Psi,
{H^\ast}^{-1}]\>,
\ena
where $H^\ast$ is defined in appendix~\ref{AppA} by eq.~\rf{ConstB}. In
this case, and according to eqs.~\rf{ConstA} and~\rf{ConstC}, the
constraints required by Proposition~1 are
\be
H^0 \in \Im = {\rm Im\/}({\rm ad\/} \Psi) \cap {\cal K}, \qquad
{H^\ast}^{-1} \in \Re = {\rm Ker\/}({\rm ad\/}
\Psi)\>,
\ee
while $h^{\geq0}=0$ is equivalent to
\be
H^0,\> {H^\ast}^{-1} \in \Re \>.
\ee
Since $\Im \subseteq \Re$, the former conditions imply the
latter. However, in general $\Im \not= \Re$ and, thus, the
constraints $h^{\geq0}=0$ are less restrictive than those required by
Proposition~1.

A particular case where Proposition~1 is satisfied without introducing
any constraint is given by the following

\begin{lem}
\label{lem2}
If the $d_1$-grade of $\Psi$ is $k=1$ and ${\cal K}^0 =
\{0\}$, Proposition~1 is satisfied without constraining the Lax
operator at all.
\end{lem}

\noindent
{\em Proof:\/} In this case, since $H^0=0$, the only constraint
would be $H^{-1} \in \Re $. However, since $[\Psi, H^{-1}]\in
{\cal K}^0 = \{0\}$, it is obviously satisfied.

\vskip 0.6truecm\noindent
This Lemma applies to all the principal hierarchies
that will be consider in the next Sections.

Lemma~1 and Proposition~1 ensure that the Lax operator $\cal L$
can be constrained such that
\be 
{\cal L} = D + q + \Psi = \hat{\Theta}_L^{-1}
\bigl(D + \Psi\bigr) \Theta_L\>,
\lab{DressPlus}
\ee
where 
\be
\Theta_L = {\rm e\/}^{\omega} {\rm e\/}^{F}\>.
\lab{ThetaL}
\ee
Consider
the group of gauge transformations given by~\rf{Gauge}. A
useful gauge fixing prescription (different from the
Drinfel'd-Sokolov gauge) is provided by the following

\begin{lem}
\label{lem3}
A consistent gauge slice can be chosen such that 
\be
\widetilde{\cal L} = D + \widetilde{q} + \Psi = \hat{\Theta}^{-1}
\bigl(D + \Psi\bigr) \Theta\>,
\lab{DressGrad}
\ee
where $\Theta= {\rm e\/}^{y}$ and $y=y(\tilde x)$ is an even
superfield taking values in ${_+{\cal A}}_{<0}$.
\end{lem}

\noindent
{\em Proof:\/} First of all, notice that the gauge 
transformations~\rf{Gauge} act on~\rf{DressPlus} according to
$\Theta_L \longrightarrow \Theta_L {\rm e\/}^{-\gamma}$.
Moreover, $\Theta_L = {\rm e\/}^{y_L}$ and $y_L \in {_+{\cal
A}}^{<0}$. Then, taking into account~\rf{Include}, one can
perform a gauge transformation generated by $\gamma_L = {\cal
P}_0(y_L)$ to construct 
\be
\Theta \>=\>  \Theta_L {\rm e\/}^{-\gamma_L} \> =\> {\rm
e\/}^{\omega} {\rm e\/}^{F}  {\rm e\/}^{-{\cal P}_0(F)} {\rm
e\/}^{-{\cal P}_0(\omega)}\>,
\lab{DosThetas}
\ee
which proves the lemma.
 
\vskip 0.6truecm

\subsection{The flow equations}

~\indent
Eq.~\rf{DressGrad} provides a one-to-one map between the
components of the (gauged fixed) Lax operator and some components of
$\Theta$ and, in our construction, the flows will be defined
as flow equations for $\Theta$. 

Let $u$ be a constant element in
${_+{\cal K}}^{\geq0}$, $D u=0$, and define the 
infinitesimal transformation~\footnote{In the following, we will
use the notation like $(\;\:)_{<0} = {\cal P}_{<0}(\;\:)$.}
\be
\delta_u \Theta = \Theta \bigl( \Theta^{-1} \hat u
\Theta\bigr)_{<0}\>.
\lab{Infinit}
\ee
Since $u\in {_+{\cal K}}^{\geq0}$, it has to be of the form
$u = e b + \epsilon \xi$, where
$b \in {\cal K}^{\geq0}\cap {\cal A}_{\overline 0}$, $\xi \in {\cal
K}^{\geq0}\cap {\cal A}_{\overline 1}$, $e\in {\cal
G}r_{\overline 0}$, and $\epsilon\in {\cal G}r_{\overline 1}$.
Then, the infinitesimal transformation can be written as 
\be
\delta_u = e D_b  + \epsilon D_\xi \>,
\lab{InfFlow}
\ee
which provides the following flow equations

\begin{defi}
\label{defi1}
The flow equations associated with $b \in {\cal
K}^{\geq0}\cap {\cal A}_{\overline 0}$ and $\xi \in {\cal
K}^{\geq0}\cap {\cal A}_{\overline 1}$ are defined by 
\be
D_b \Theta = \Theta \bigl( \Theta^{-1} b
\Theta\bigr)_{<0}\>, \qquad
D_\xi \Theta = - \hat\Theta \bigl( {\hat\Theta}^{-1} \xi
\Theta\bigr)_{<0}\>,
\lab{Flows}
\ee
where $D_b$ and $D_\xi$ are bosonic and Grassmann
derivatives, respectively.
\end{defi}

These flows do not commute among themselves. Instead, they close
the following superalgebra.

\begin{theo}
\label{theo1}
For any $b, c \in {\cal K}^{\geq0}$,
\be
[D_b , D_c] = D_{[c,b]}\>.
\lab{CommD}\ee
\end{theo}

\noindent
{\em Proof:\/} Let us consider the infinitesimal transformations
corresponding to two elements $u,v$ in ${_+{\cal K}}^{\geq0}$. Then,
\be
\delta_u \delta_v \Theta =\Theta \biggl( \bigl(\Theta^{-1} \hat u
\Theta\bigr)_{<0} \bigl(\Theta^{-1} \hat v
\Theta\bigr)_{<0} + [\Theta^{-1} \hat v
\Theta, (\Theta^{-1} \hat u
\Theta\bigr)_{<0}]_{<0} \biggr)\>,
\ee
and it is straightforward to check that
\be 
[\delta_u , \delta_v] \Theta = -\Theta \bigl(\Theta^{-1} [\hat u,
\hat v] \Theta\bigr)_{<0} = \delta_{[v,u]}\>.
\ee
Then, the proof of the theorem can be completed by considering the
relation between the infinitesimal transformations $\delta_u$ and
$\delta_v$ and the flow equations given by~\rf{InfFlow}.

\vskip .6truecm
The flow equations~\rf{Flows} can be expressed in
terms of Lax operators. Let $u \in {_+{\cal K}}^{\geq0}$ and
consider
\be
\Delta_u = \delta_u - \bigl( \Theta^{-1} \hat
u\Theta\bigr)_{\geq0} =  \Theta^{-1}\bigl( \delta_u -  \hat
u\bigr)\Theta\>,
\lab{Delta}
\ee
where we have used the notation $\delta_u \Theta = (\delta_u
\Theta) + \Theta \delta_u$ and eq.~\rf{Infinit}.
Then, if $u = e b + \epsilon \xi$, the Lax operators corresponding
to $b$ and $\xi$ can be defined by analogy with~\rf{InfFlow}:
\be
\Delta_u = e {\cal L}_b  + \epsilon {\cal L}_\xi \>.
\lab{InfLax}
\ee
In other words,

\begin{defi}
\label{defi2}
Let $b \in {\cal K}^{\geq0}\cap {\cal
A}_{\overline 0}$ and $\xi \in {\cal K}^{\geq0}\cap {\cal
A}_{\overline 1}$, the (gauge fixed) Lax operators associated with
$D_b$ and $D_\xi$ are defined by
\bea
& &{\cal L}_b = D_b - \bigl( \Theta^{-1} b \Theta\bigr)_{\geq0}
= \Theta^{-1}\bigl( D_b - b \bigr)\Theta\>, \nn
& &{\cal L}_\xi = D_\xi + \bigl( {\hat\Theta}^{-1} \xi
\Theta\bigr)_{\geq0}
= {\hat\Theta}^{-1}\bigl(D_\xi + \xi \bigr) \Theta\>,
\lab{Laxes}
\ena
where $D_b \Theta = (D_b \Theta) + \Theta D_b$ and $D_\xi \Theta =
(D_\xi \Theta) + \hat \Theta D_\xi$. The Lax operators ${\cal L}_b$
and ${\cal L}_\xi$ are even and odd, respectively.
\end{defi}

Then, Theorem~1 translates into the following (super) commutation
relations

\begin{theo}
\label{theo2}
For each $u,v \in {_+{\cal K}}^{\geq0}$, $b,c \in {\cal
K}^{\geq0}\cap {\cal A}_{\overline 0}$, and $\xi,\eta \in {\cal
K}^{\geq0}\cap {\cal A}_{\overline 1}$, the following commutation
relations are satisfied
\bea
& &[\Delta_u, \Delta_v] = \Delta_{[v,u]}\>,\qquad 
[{\cal L}_b, {\cal L}_c] = {\cal L}_{[c,b]}\>, \nn
& &[{\hat {\cal L}}_b, {\cal L}_\xi] = - [{\cal L}_\xi, {\cal L}_b]
= {\cal L}_{[\xi, b]} \>, \nn
& & [{\hat {\cal L}}_\xi, {\cal L}_\eta] = [{\hat {\cal L}}_\eta,
{\cal L}_\xi] = {\cal L}_{[\eta, \xi]}\>.
\lab{Commut}
\ena
\end{theo}

\noindent
{\em Proof:\/} It is straightforward by using~\rf{Delta},
\rf{Laxes}, and~\rf{CommD}. For example,
\bea
[\Delta_u, \Delta_v] &=& \Theta^{-1} [\delta_u -\hat u, \delta_v
-\hat v] \Theta = \Theta^{-1}\Bigl( \delta_{[v,u]} +[\hat u, \hat v]
\Bigr)\Theta  \nn
&=& \Theta^{-1}\Bigl( \delta_{[v,u]} -[\hat v, \hat u]
\Bigr)\Theta = \Delta_{[v,u]}\>.
\ena

\vskip .6truecm

\subsection{Supersymmetric flow equations}

~\indent
One of the main motivations of the work of Delduc and
Gallot in~\cite{DelGal} was to obtain supersymmetric hierarchies.
In other words, hierarchies whose flow equations are compatible
with a supersymmetry transformation relating commuting to
anticommuting fields. 

Since both the construction of~\cite{DelGal}
and ours are formulated directly in $N=1$ superspace with
coordinates $\tilde x=(x, \theta)$, there is a built-in
supersymmetry transformation induced by the
covariant derivative: $\delta_{\eta}^{\rm SUSY} q =
\eta Dq$, where $\eta$ is any constant Grassmann-odd parameter. By
construction, and according to eqs.~\rf{Laxes}, \rf{DressGrad},
\rf{Commut}, and~\rf{EvenLax}
\be
{\cal L}_\Psi = \widetilde{\cal L}\>, \quad {\rm and}\quad
{\cal L}_\Lambda = {1\over2}
[{\hat {\cal L}}_\Psi, {\cal L}_\Psi] = \widetilde{L}\>,
\lab{IdLax}
\ee
which implies the identities
\be
D_\Psi = D  \quad {\rm and}\quad D_\Lambda = {\partial\over
\partial x}\>.
\lab{IdentD}
\ee
All this allows one to extend the supersymmetry transformation to
the space of functionals of $\Theta$ or, equivalently, to the
space of gauge invariant functionals of $q$ by means of
$\delta_{\eta}^{\rm SUSY}  = \eta D_\Psi$. Therefore, a flow
equation will be supersymmetric if, and only if, it commutes with
$D_\Psi$.

Taking into account the commutation relations~\rf{CommD}, it is
straightforward to prove 
 
\begin{theo}
\label{theo3}
The flow equation $D_\xi$ is supersymmetric if, and only
if, 
\be
\xi \in \Re^{\geq0} = {\rm Ker\/}({\rm ad\/} \Psi)^{\geq0}\>.
\ee
\end{theo}

Remarkably, although they are generally non-local, the supersymmetric
flows can be written in terms of the same (super) classical r-matrix
$\cal R$ used in eq.~\rf{LocFlow}.

\begin{Prop}
\label{Prop2}
For each $b \in \Re^{\geq0}\cap {\cal
A}_{\overline 0}$ and $\xi \in \Re^{\geq0}\cap {\cal
A}_{\overline 1}$, the
corresponding supersymmetric flow equations on $q$ can be written
as
\be
D_b q = [{\cal R}\bigl({\hat \Theta}_L^{-1} b {\hat
\Theta}_L\bigr), {\cal L}]\>, \qquad
D_\xi q = [{\cal R}\bigl(\Theta_L^{-1} \xi {\hat
\Theta}_L\bigr), {\cal L}]\>,
\lab{SusyFlow}
\ee
where ${\cal R} = {1\over2}({\cal P}_{\geq0} - {\cal P}_{<0})$ and 
$\Theta_L$ is given by~\rf{ThetaL}.
\end{Prop}

\noindent
{\em Proof:\/}
Consider $u= eb + \epsilon\xi \in {_+\Re}^{\geq0}$. Since $[u,\Psi]=0$,
and taking into account~\rf{IdLax},  the infinitesimal transformation of
the gauged fixed
$\widetilde{q}$ can be written as $[{\hat \Delta}_u,
\widetilde{\cal L}] =0$, where $\Delta_u$ has been defined
in~\rf{Delta}. Moreover, according to~\rf{DosThetas}
and~\rf{Infinit},
\bea
\Delta_u &=& {\rm e\/}^{\gamma_L} \Theta_L^{-1} \bigl(\delta_u -\hat
u\bigr)\Theta_L {\rm e\/}^{-\gamma_L} \nn
&=& {\rm e\/}^{\gamma_L} \Bigl( \delta_u - (\Theta_L^{-1} \hat u
\Theta_L)_{\geq0} + \eta_L \Bigr) {\rm e\/}^{-\gamma_L}\>,
\ena
where 
\be
\eta_L = {\rm e\/}^{-\gamma_L} (\delta_u {\rm
e\/}^{\gamma_L}) \in {_+{\cal A}}_0^{<0}\>.
\ee
Therefore, according to~\rf{DressGrad}, the infinitesimal
transformation of $q$ is given by
\be
[{\rm e\/}^{-\hat\gamma_L} {\hat \Delta}_u {\rm
e\/}^{\hat\gamma_L}, {\cal L}] =0\>.
\ee
This leads to
\be
\delta_u q = [\bigl({\hat \Theta}_L^{-1} u
{\hat\Theta}_L\bigr)_{\geq0} -
\eta_L, {\cal L}]\>, 
\ee
which is equivalent to~\rf{SusyFlow} up to an infinitesimal gauge
transformation generated by $\eta_L$ that can be omitted because
flow equations are defined on the space of gauge invariant
functionals of $q$. Finally, the proof of the Proposition follows
by considering~\rf{DressPlus}, \rf{InfFlow}, and taking into
account that $[u, \Psi]=0$.

\vskip 0.6truecm
In particular, Proposition~2 allows one to check that the flow
equations~\rf{LocFlow} considered by Delduc and Gallot are recovered
from our flow equations with the constant elements in $\cal
Z$, the centre of ${\cal K}={\rm Ker\/}({\rm ad\/}\Lambda)$.

\begin{cor}
\label{cor1}
If $b\in {\cal Z}^{\geq0}$, $D_b$  coincides with
$\partial/\partial t_b$ in eq.~\rf{LocFlow}.
\end{cor}

\noindent
{\em Proof:\/} It is straightforward by considering Proposition~2
and~\rf{ThetaL}:
\be
{\hat \Theta}_L^{-1} b {\hat\Theta}_L = {\rm e\/}^{-\hat F} {\rm
e\/}^{- \hat \omega} b {\rm e\/}^{\hat \omega} {\rm e\/}^{\hat F}
= {\rm e\/}^{-\hat F}  b {\rm e\/}^{\hat F} = B_b\>,
\ee
where we have used that $[b, \omega]=0$.

\vskip .6truecm
Taking into account this result and Lemma~1, the flows
associated with the elements in $\cal Z$ are local flows in the
sense that $D_b q$ is a super-differential polynomial of $q$.
In general, the flows associated with the Lax operators~\rf{Laxes}
are non-local. However, it is important to notice that, {\em a
priori\/}, nothing prevents the existence of other local flows
than those associated with $\cal Z$. In particular, there could
be Grassmann-odd local flows. An example is provided by the the
following Lemma.

\begin{lem}
\label{lem4}
If there exists an element $\xi\in {\cal K}^{\geq0}$ such
that
$[\xi, {\cal K}^{<0}] \subset {\cal K}^{<0}$, the flow equation
associated with $D_{\xi}$ is local.
\end{lem}

\noindent
{\em Proof:} Without loss of generality, let us suppose that
$\xi$ is odd. Then, $D_{\xi}$ is induced by
the Lax operator
\be
{\cal L}_{\xi} = D_{\xi} + \bigl({\hat \Theta}^{-1} \xi
\Theta \bigr)_{\geq0} =
D_{\xi} + \bigl({\rm e\/}^{-\hat F} {\rm e\/}^{-\hat \omega}\xi
{\rm e\/}^{ \omega} {\rm e\/}^{ F}\bigr)_{\geq0} \>.
\ee
However, since ${\cal A}^{<0}\subset {\cal A}_{\leq0}$, in this case
\be
\bigl( {\rm e\/}^{-\hat \omega}\xi {\rm e\/}^{ \omega} \bigr)_{\geq0} 
= \xi + \bigl( {\rm e\/}^{-\hat \omega}\xi {\rm e\/}^{ \omega}
\bigr)_{0}^{<0}\>,
\ee
and, hence,
\be
{\cal L}_{\xi} = D_{\xi} + \bigl({\rm e\/}^{-\hat F} \xi {\rm e\/}^{
F}\bigr)_{\geq0} + \sigma\>, 
\ee
where $\sigma \in {\cal A}_{0}^{<0}$. Acting on $q$, the only
effect of $\sigma$ is to generate an infinitesimal gauge
transformation. Therefore, on the space of gauge invariant
functionals the flow $D_\xi$ can be equivalently defined by 
${\cal L}_{\xi} -\sigma$, which proves that $D_{\xi}$ indeed
is local.

\vskip0.6truecm\noindent
Examples where this lemma applies will be presented in Section
\ref{sec-5}. 

\vskip0.6truecm

\sect{Local and non-local conserved quantities}
\label{sec4}

~\indent
In this section, we will construct an infinite number of local and
non-local conserved quantities with respect to the bosonic flow
equations~\rf{LocFlow} associated with $\cal Z$, the centre of
${\cal K}={\rm Ker\/}({\rm ad\/} \Lambda)$. We will also investigate
which of them are supersymmetric.

Since most of the following expressions involve non-local terms, it is
necessary to specify the boundary conditions. We will restrict
ourselves to hierarchies where the superfield $q=q(\tilde x)$ goes
rapidly to zero at $x= \pm\infty$. Then, since $F$ and $H$ in Lemma~1 
are super-differential polynomials of $q$, they also vanish at $x=
\pm\infty$:
\be
q \Bigm|_{x= \pm\infty} = F \Bigm|_{x=
\pm\infty} = H \Bigm|_{x= \pm\infty} =0\>.
\lab{Boundary}
\ee

Consider the flow equation associated with
any $b\in {\cal Z}^{\geq0}$. According to~\rf{ThetaL}
and~\rf{DosThetas},
\bea
D_b\Theta = \Theta \bigl(\Theta^{-1} b\Theta \bigr)_{<0} 
&=& {\rm e\/}^{\omega} {\rm e\/}^{F} \bigl({\rm e\/}^{-F} {\rm
e\/}^{-\omega} b {\rm e\/}^{\omega} {\rm e\/}^{F} \bigr)_{<0} {\rm
e\/}^{\gamma_L} \nn
&=& {\rm e\/}^{\omega} {\rm e\/}^{F} \bigl({\rm e\/}^{-F}  b  {\rm
e\/}^{F} \bigr)_{<0} {\rm e\/}^{\gamma_L}\>,
\lab{SinOmega}
\ena
where it has been taken into account that $\gamma_L \in {\cal
A}_0$ and $[b, \omega]=0$. Using~\rf{Boundary}, this leads to
\be
D_b\Theta \Bigm|_{x= \pm\infty} = {\rm e\/}^{\omega}
(  b )_{<0}
{\rm e\/}^{\gamma_L}\Bigm|_{x= \pm\infty} =0\>,
\lab{ConservaA} 
\ee
where we have used that $b\in {\cal Z}^{\geq0}\subset {\cal
A}_{\geq0}$. All this allows one to prove
the following

\begin{Prop}
\label{Prop3}
All the components of $\omega \mid_{x= \pm\infty} $
are conserved quantities with respect to the (bosonic) flows
associated with ${\cal Z}^{\geq0}$.
\end{Prop}

\noindent
{\em Proof:} First of all, using~\rf{Boundary} and~\rf{DosThetas},
\be
\Theta\Bigm|_{x= \pm\infty} = {\rm e\/}^{\omega} {\rm
e\/}^{-{\cal P}_0(\omega)}\Bigm|_{x= \pm\infty}\>.
\ee
Consider $\omega= \sum_{j>0} \omega^{(j)} \xi^{-j}$, where
$\xi^{-j} \in {\cal K}^{-j}$ and $\omega^{(j)}$ is a superfield
taking values in the Grassmann algebra ${\cal G}r$. Then,
decompose the conservation equation~\rf{ConservaA} in its graded
components with respect to~$d_1$. The component whose grade
equals~$-1$ reads
\be
D_b \omega^{(1)}\Bigm|_{x= \pm\infty} \bigl(\xi^{-1} - {\cal
P}_0(\xi^{-1}) \bigr) =0\>,
\ee
which implies that $\omega^{(1)}|_{x= \pm\infty}$ is
conserved. Next, let us suppose that $\omega^{(j)}|_{x=
\pm\infty}$ is conserved for all $1\leq j<N$ and consider the
grade-$N$ component of~\rf{ConservaA},
\be
D_b \biggl(\omega^{(N)} \bigl(\xi^{-N} -
{\cal P}_0(\xi^{-N})\bigr) \> +\> \cdots \biggr)\Bigm|_{x=
\pm\infty} =0\>,
\ee
where the ellipsis indicates products of $\omega^{(j)}|_{x=
\pm\infty}$'s with $j<N$. Therefore, one concludes that
$\omega^{(N)}|_{x= \pm\infty}$ is also conserved, which
completes the proof.

\vskip0.6truecm
Notice that the components of $\omega$ are non-local functions of
$q$; in other words, they involve $D^{-1}q$ and, hence, are not super
differential polynomials of $q$. This is the reason why their asymptotic
values do not vanish in spite of the boundary conditions~\rf{Boundary}.
However, some of the components of $\omega|_{x= \pm\infty}$ provide
the local conservation laws given by~\rf{ConsL}.

\begin{Prop}
\label{Prop4}
For each $\xi\in [{\cal K}, {\cal K}]^\bot$,
\be
J_\xi^L= \int_{-\infty}^{+\infty} dx \> \partial_x \langle\xi,
\omega\rangle = \int d{\tilde x}\> \langle\xi, H\rangle
\lab{ConservaLoc}
\ee
is a supersymmetric local conserved quantity.
\end{Prop}

\noindent
{\em Proof:} Consider~\rf{AbelPlus}; since $\xi\in [{\cal K},
{\cal K}]^\bot$,  it leads to
\be
\langle\xi, H\rangle\> = \langle\xi, {\rm e\/}^{-\hat \omega}
(D{\rm e\/}^{ \omega}) + {\rm e\/}^{-\hat \omega}\Psi {\rm
e\/}^{\omega} -\Psi \rangle\> = \langle\xi, D \omega\rangle\>.
\ee
Then, taking into account $\partial_x = D^2$ and~\rf{Boundary},
\be
J_\xi^L= \int_{-\infty}^{+\infty} dx \> D\langle\xi, H\rangle\> =
\int_{-\infty}^{+\infty} dx \> {\partial\over
\partial\theta}\langle\xi, H\rangle\> = \int d{\tilde x}\>
\langle\xi, H\rangle\>.
\lab{Dtod}
\ee
Being written as the superintegral of a super differential
polynomial of the superfield~$q$, the local conserved quantity
$J_\xi^L$ is manifestly invariant with respect to the supersymmetry
transformation $\delta_{\eta}^{\rm SUSY} q= \eta Dq$, where $\eta$ is
any constant Grassmann-odd parameter. 

\vskip0.6truecm
A convenient definition of the non-local conserved quantity associated
with a generic element $\chi \in {\cal K}^{\geq0}$ is
\be
J_\chi =\int_{-\infty}^{+\infty} dx\> 
\partial_x\>  {\rm str\/}\> \bigl(\chi {\rm
e\/}^{\omega}\bigr)\>,
\lab{Jchi}
\ee
where `str'  is the supertrace (see appendix~\ref{AppB}).
Taking into account proposition~\ref{Prop3}, $J_\chi$ is obviously
conserved. Let us calculate the supersymmetric transformation of $J_\chi$.
First of all, recall that~\rf{IdentD} allows one to extend the
supersymmetry transformation $\delta_{\eta}^{\rm SUSY}$ to the space of
functionals of $\Theta$. Therefore,
\be
\delta_{\eta}^{\rm SUSY} J_\chi = \eta D_\Psi J_\chi =
\eta\> \int_{-\infty}^{+\infty} dx\> 
\partial_x\>  {\rm str\/}\> \bigl(\chi \> D\> {\rm
e\/}^{\omega}\bigr) \>,
\ee
where, using~\rf{AbelPlus}, 
\be
D{\rm e\/}^{\omega} = {\rm e\/}^{\hat\omega} H + {\rm
e\/}^{\hat \omega}\Psi - \Psi {\rm e\/}^{\omega}\>.
\lab{ConservaC}
\ee
Moreover, it is easy to show that
\be
{\rm str\/}\> \Bigl(\chi\> \bigl( {\rm e\/}^{\hat \omega}\Psi - \Psi
{\rm e\/}^{\omega}\bigr) \Bigr) = - {\rm str\/}\> \bigl([\chi, \Psi]
{\rm e\/}^{\omega} \bigr)\>,
\lab{Strmagic}
\ee
which allows one to write the supersymmetric transformation of $J_\chi$
as follows 
\be 
\delta_{\eta}^{\rm SUSY} J_\chi = \eta\> \int_{-\infty}^{+\infty}
dx \> \partial_x \>{\rm str\/}\> \bigl(\chi \> {\rm e\/}^{\hat\omega}
\>H\bigr) 
- \eta\> \int_{-\infty}^{+\infty}
dx \> \partial_x \>{\rm str\/}\> \bigl([\chi, \Psi] \> {\rm
e\/}^{\hat\omega}\bigr) \>.
\lab{LocNloc}
\ee

At this stage, we would like to make a comment about
partial integrations in expressions with non-local terms. It is
clearly `not' true that $\int dx\> \partial_x P(x) =0$ if $P(x)$ is a
generic non-local expression; for example, if $P(x) =
\partial_{x}^{-1} q$ the statement is obviously false. If, in
agreement with~\rf{Boundary}, we choose boundary conditions such that
all local expressions vanish sufficiently fast at $x=\pm \infty$ and
we take $P(x) = L(x) N(x)$, where $L(x)$ is a local expression while
$N(x)$ may be non-local, then the statement is true. This is clear
since
\be
\int_{-\infty}^{+\infty} dx \partial_x P(x) = \bigl( L(x) N(x) \bigr)
\Big|^{+\infty}_{-\infty} =0
\lab{partial-int}
\ee
due to the vanishing of $L(x)$ at $x=\pm \infty$. So, as long as
we limit ourselves to expressions of the form
`local'$\times$`non-local', the use of partial integration is
justified.

Taking into account all this, and since $H$ is local, the first term
on the right-hand side of~\rf{LocNloc} vanishes and, therefore, 
\be 
\delta_{\eta}^{\rm SUSY} J_\chi = -\eta J_{[\chi, \Psi]}\>,
\lab{NLSusy}
\ee
which allows one to single out the set of supersymmetric non-local
quantities.

\begin{Prop}
\label{Prop5}For each $\chi \in \Re^{\geq0} = {\rm Ker\/}({\rm ad\/}
\Psi)^{\geq0}$ the non-local conserved quantity
\be
J_\chi =\int_{-\infty}^{+\infty} dx\> \partial_x\>  {\rm
str\/}\> \bigl(\chi \> {\rm e\/}^{\omega}\bigr) = 
\int d\tilde x\> {\rm str\/}\> \bigl(\chi \>
{\rm e\/}^{\hat\omega} \>H\bigr)
\lab{ConservaNL}
\ee
is supersymmetric with respect to $\delta_{\eta}^{\rm SUSY} q= \eta
Dq$.
\end{Prop}

\noindent
The proof follows directly from~\rf{NLSusy}, and the expression of
$J_\chi$ as a superintegral is obtained by using
$\partial_x =D^2$ and~\rf{ConservaC} (see also eq.~\rf{Dtod}). 

\vskip 0.6truecm

For a general $\chi$, the conserved quantity $J_\chi$ is
a function of $\theta$ and, therefore, has two scalar components.
However, when $[\chi, \Psi]=0$ the right-hand side
of~\rf{ConservaNL} shows that the component along $\theta$ vanishes.
In order to split the two scalar components of a generic conserved
quantity let us consider $\partial_x= D^2$ in eq.~\rf{Jchi}
together with eqs.~\rf{ConservaC} and~\rf{Strmagic}:
\be
J_\chi= \int_{-\infty}^{+\infty} dx\> 
D\>  {\rm str\/}\> \bigl(\chi {\rm
e\/}^{\hat \omega}H - [\chi, \Psi] {\rm
e\/}^{\omega}\bigr) \>.
\ee
Then, using the definition of the superderivative $D$ and taking into
account that $H$ is local, $J_\chi$ can be written as
\be
J_\chi = I_\chi - \theta\> I_{[\chi, \Psi]}\>,
\lab{Split}
\ee
where the scalar components are given by 
\be
I_\chi =  \int d\tilde x\> D\> {\rm str\/}\> \bigl(\chi \>
{\rm e\/}^{\omega}\bigr) =
\int d\tilde x\> {\rm str\/}\> \bigl(\chi \>
{\rm e\/}^{\hat\omega} \>H - [\chi, \Psi] {\rm e\/}^{\omega} \bigr)\>,
\lab{ConservaNLplus}
\ee
which exhibits that $J_\chi =I_\chi$ whenever $[\chi,
\Psi]=0$, in agreement with~\rf{ConservaNL}. 

If~$j$ and~$k$ are the
$d_1$-grades of $\chi$ and $\Psi$, respectively, it is
straightforward to prove that $I_\chi$ has dimension $j/2k$ with
respect to a scale~\label{scale} transformation where the dimension of $x$
and $\theta$ is~1 and~$1\over2$. Moreover, $I_\chi$ is Grassmann-odd or
Grassmann-even depending on whether $\chi$ is a odd or even element of
the superalgebra, respectively.

Finally, it is worth noticing that, according to theorem~\ref{theo3}, the
non-local supersymmetric conserved quantities are in one-to-one relation
with the supersymmetric flow equations. However, we have not been able
to find a deeper relationship between them where, for example, $J_\chi$
could be understood as the generator of the flow $D_\chi$.  

\sect{Principal Hierarchies} 
\label{sec-5}

In this section we will restrict ourselves to consider supersymmetric
KdV systems in which $\Psi$ is a {\em principal} fermionic element of
$\ca$, {\it i.e.\/}, 
$$
\Psi = \sum_i a_i E_{\a_i}
$$
where the sum is over all the simple roots of the algebra, and the $a_i$
are non-zero numbers. A necessary condition for the existence of
such an element is that $\ca$ is equipped with a fermionic simple root
system, and in table \ref{FSRS} we give a complete list of all such
affine Lie superalgebras; for future reference the table contains also
the bosonic subalgebras. 
\begin{table}
\begin{center}
\begin{tabular}{|r|r|}
\hline 
Superalgebra & Bosonic subalgebra\\ \hline 
$A(m,m)^{(1)}$ &  $A_m^{(1)} \oplus A_m^{(1)}$ \\
$A(2m,2m)^{(4)}$ & $A_{2m}^{(2)} \oplus A_{2m}^{(2)}$ \\
$A(2m{-}1,2m{-}1)^{(2)}$ & $A_{2m{-}1}^{(2)} \oplus A_{2m{-}1}^{(2)}$ \\
$A(2m{-}2,2m{-}1)^{(2)}$ & 
$A_{2m{-}2}^{(2)} \oplus A_{2m{-}1}^{(2)} \oplus U(1)^{(1)}$ \\ 
$B(m,m)^{(1)}$ & $B_m^{(1)} \oplus C_m^{(1)}$ \\
$D(m{+}1,m)^{(1)}$ & $D_{m{+}1}^{(1)} \oplus C_{m}^{(1)}$ \\
$D(m,m)^{(2)}$ & $D_m^{(2)} \oplus C_m^{(1)}$ \\
$D(2,1;\a)^{(1)}$ & $ A_1^{(1)} \oplus A_1^{(1)} \oplus A_1^{(1)} $
\\
\hline 
\end{tabular}
\end{center}
\caption{Affine Lie superalgebras with fermionic simple root
systems. In all cases $m\ge 1$. Note that, for low values of $m$, the
even subalgebras may fit into the general scheme only modulo 
standard identifications:
$A_0=\{0\},\,A_1^{(2)}\simeq A_1^{(1)},\,
B_1\simeq C_1\simeq A_1,\,
D_1\simeq U(1)$ and $D_2\simeq A_1\oplus A_1$.}
\label{FSRS}
\end{table}

However, we will start by giving two lemmas which apply to the general
case, namely: 

\begin{lem}
\label{lemm1}
An element $\L\in\ca_\bz$ is semi-simple in $\ca$ if and only if
it is semi-simple in $\ca_\bz$.
\end{lem}
It is well known that an element in a Lie algebra 
is semi-simple if and only if it can be taken to be element of a
Cartan subalgebra. The same argument can be immediately applied for Lie
superalgebras, and 
then the lemma follows, since a Cartan subalgebra of $\ca_\bz$ is a
Cartan subalgebra of $\ca$. 

\begin{lem}
\label{lemm2} 
If $\L\in\ca$ is regular in $\ca_\bz$, then 
$[\ck_\bo,\ck_\bo] \subset \cz$, where $\ck=\ker(\ad(\L))$ and $\cz$
is the centre of $\ck$. 
\end{lem}

$\L$ is regular in $\ca_\bz$, so we can take a basis in which
$\ck_\bz$ is the Cartan subalgebra $H$ of
$\ca$. $\ck_\bo$ consists of odd root-generators
$E_\a$ with eigenvalue zero under $\ad(\L)$. 
If $E_\a,\,E_\b \in \ck_\bo$, then $\a\cdot\b=0$,  
%
%
({\em including} the case $\a=\b$), namely: 
\begin{itemize}
\item[1)] $\b=\a$: $\a\cdot\a\not=0$ implies that $2\a$ is a root (see
\cite{Liesuper}) which is a contradiction, since 
$[K_\bo,K_\bo]\subset H$. 
\item[2)] $\b\not=\a$: $\a\cdot\b\not=0$ implies that either $\a+\b$ or
$\a-\b$ is a root (this can be easily shown using the Jacobi identity
on the generators $E_\a,\,E_\b$  and $E_{-\b}$). This is again a
contradiction, for the same reason as above. 
\end{itemize}
This implies that if $h_\a\in [\ck_\bo,\ck_\bo]$ and 
$e_\b\in \ck_\bo$, then $[h_\a,E_\b]=(\a\cdot\b) E_\b=0$, and
therefore $h_\a\in \cz$. 

For future reference, we will formulate part of this lemma as a
corollary:  
\begin{cor} \label{cor2} For $\L$ regular in $\ca_\bz$ and in a
basis of root-generators for $\ck_\bo$: If $\a$ and $\b$ are roots
such that $E_\a,\,E_\b\in\ck_\bo$, then $\a^2=\b^2=\a\cdot\b=0$.  
\end{cor}

Defining $[\ck_\bo,\ck_\bo]\equiv \cz' \subset \cz$, an
immediate consequence of the arguments given in the proof of the lemma
is the following  
\begin{cor} \label{coro1} $\cz'$ is completely degenerate\end{cor}
since for any two elements $h_\a,\,h_\b\in \cz'$ we have 
$\langle h_\a,h_\b\rangle = \a\cdot\b=0$.

\subsection{Principal Elements}
\label{prin} 

Let now $\Psi$ be a principal element of $\ca$, where $\ca$ is one of
the affine Lie superalgebras listed in table \ref{FSRS} except
$D(2,1;\a)$; we have
chosen not to consider this exceptional superalgebra in detail in
order to keep 
the exposition as simple as possible, but it can be treated using the
same methods as the other algebras. 
We write a grading $d$ in the form $(s_0,s_1,\ldots,s_r)$, where $s_i$
is the grade of the generator corresponding to the simple root
$\a_i$. Then, we call principal hierarchies those associated with a
principal element $\Psi$, the principal grading $d_1= (1,1,\ldots,1)$,
and another gradation $d_0$. It is convenient to distinguish the
following cases: 
%
\begin{itemize}
\item[--] $d_0 = (1,0,0,\ldots,0)$ is the homogeneous grading. This
corresponds to what we will call SKdV type hierarchies. 
\item[--] $d_0 = d_1$ are both the principal grading. We will 
call the corresponding hierarchies SmKdV hierarchies, supersymmetric
modified KdV hierarchies. 
\item[--] $d_0$ is in
between~\footnote{It is possible to make this 
precise by defining a partial ordering of gradings, see 
\cite{miramontesetal}} the
homogeneous and the principal gradings. The resulting hierarchies are
the SpmKdV hierarchies, 
supersymmetric partially modified KdV hierarchies. 
\end{itemize} 

The choice of $d_0$ is always restricted by the non-degeneracy
condition \rf{Degen}. When $\Psi$ is principal, this condition is
satisfied both for the SKdV and SmKdV type hierarchies, except for the
algebra $A(m,m)^{(1)}$, where the choice 
$d_0 = (1,0,0,\ldots,0)$ does not satisfy \rf{Degen}. Instead we
can consider a 
`minimally modified' SpmKdV hierarchy; we can for example 
follow \cite{DelGal} and take $d_0 = (1,0,0,\cdots,0,0,1)$. 

In the principal hierarchies, the $d_1$- (principal) grade of $\Psi$ is
by definition 1. In addition, $\ck^0 =
\{0\}$, which follows directly from the corresponding statement for
bosonic  hierarchies, since $\ck^0$ is bosonic. 
This implies that lemma \ref{lem2} applies, {\it i.e.\/}, it is always
possible, without constraining the Lax operator, to find an even
superfield $\Theta_L$ such that 
$\hat{\Theta}_L \cl \Theta_L^{-1} = D + \Psi$. We can therefore
immediately use
sections \ref{sec3} and \ref{sec4} to define flow equations and
conserved quantities. 

Since all
bosonic simple roots of $\ca_\bz$ can be obtained as sums of two
simple roots of $\ca$, $\L = \hf [\Psi,\Psi]$ is a principal element
in $\ca_\bz$. $\L$ is therefore regular in $\ca_\bz$, and
according to lemma \ref{lemm1} regular in $\ca$. 
According to table \ref{FSRS}, $\ca_\bz$ is always of the form 
$\cg_1\oplus \cg_2$, or $\cg_1\oplus \cg_2\oplus U(1)^{(1)}$ in the
case of $A(2m{-}2,2m{-}1)^{(2)}$. The subalgebra $\cg_1$ is always simple
except in the case of $D(2,1)^{(1)}$ where $\cg_1 = D_{2}^{(1)}\simeq
A_{1}^{(1)} \oplus A_{1}^{(1)}$.  Even in the case of
$A(2m{-}2,2m{-}1)^{(2)}$, $\L$ is (with one exception) an element in
$\cg_1\oplus\cg_2$. This follows from the fact that $\L$ has grade 2,
while the $U(1)^{(1)}$-part of $A(2m{-}2,2m{-}1)^{(2)}$ has grade
larger than 2. The only exception is $A(0,1)^{(2)}$, 
where the even subalgebra is $U(1)^{(1)}\oplus A_{1}^{(1)}$ and where
the $U(1)^{(1)}$ subalgebra has grade 2; we will treat this example in
detail in section \ref{SKdV}. It is therefore clear that we can write
$\L = \L_1 + \L_2$ where $\L_i$ is a principal element in $\cg_i$. 

The problem of finding the even subalgebra $\ck_\bz$ of $\ck$, and of
finding explicit realizations of its elements, has therefore been
reduced to the problem of finding the kernel of $\ad(\L)$ in the case
where $\L$ is a principal element in a (bosonic) Lie algebra $\cg$. This
question has been studied in e.g. \cite{DrSo}. The elements of
$\ker(\ad(\L))$ are in one-to-one 
correspondence with the set of exponents $\{e_i\}$ of $\cg$, and are
essentially of the form $\L^{e_i}$; for more details and for the
exceptions see appendix \ref{App} and \cite{DrSo}. 

The task which is left is to find the odd subspace $\ck_\bo$ of
$\ck$. In order to do this, consider 
the covering homomorphism $\phi$, the mapping which sends the affine 
algebra ${\cal A} = {\cal L}({\cal G}, \tau) \subset {\cal G}\otimes 
\fl{C}[\lambda, \lambda^{-1}]$ into the corresponding finite algebra 
$\cg$ by setting $\l$ equal to 1 (or any non-zero number). 
Take a root-basis for $\cg$ such that $\phi(\ck_\bz)$ is the Cartan
subalgebra. As shown in corollary \ref{cor2}, any two (fermionic)
roots $\a$ and $\b$ such that  
$E_\a,\,E_\b\in\phi(\ck_\bo)$ must satisfy $\a^2=\b^2=\a\cdot\b=0$. 
The space spanned by those roots $\a$ such
that $E_\a\in\phi(\ck_\bo)$ is therefore completely degenerate. Denote
by $(d_+,d_-)$ the signature of the root-space of $\ca$~\footnote{In
the case of $A(m,n)$ we actually use the signature of a space which
includes the root-space as a subspace, see page~\pageref{VV}}; the
maximum dimension of a completely degenerate subspace is min($d_+,d_-$). 
Thus the maximal dimension of $\phi(\ck_\bo)$ is 
$2\,\mbox{min}(d_+,d_-)$ (for each root $\a$ we have $E_{\pm\a}$). 
\label{Dim}

In appendix \ref{App} we show that $\Psi$ together with
the elements of $[\Psi,\ck_\bz]$, $[[\Psi,\ck_\bz],\ck_\bz]$, etc.,
saturate this limit, except in the case of  the algebra
$A(2m{-}1,2m{-}1)^{(2)}$. In this last case, the bound allows two extra
sets of elements of $\ck_\bo$ which we find by explicit calculations, see
eq.~\rf{xi}. 

The elements of $\ck_\bo$ turn out to fall naturally into two distinct 
classes of elements (four in the case of $A(2m{-}1,2m{-}1)^{(2)}$). We
denote the elements in these two (four) classes by $\psi_i$ and
$\bar{\psi}_i$ (plus $\xi_i$ and $\bar{\xi}_i$) respectively, where $i$
is the principal grade. A convenient way to parametrize the grades is to
write $\psi_{-1+2i}$ and $\bar{\psi}_{1+2i}$,  where for $A(m,m)^{(1)}$
$i\in\fl{Z}$, and for all other algebras $i\in 2\fl{Z}+1$. We identify
$\Psi=\psi_1$. \label{ferm-elts} The elements $\xi_i$ and $\bar{\xi}_i$
of $A(2m{-}1,2m{-}1)^{(2)}$ have grades $2m{-}1$ and $6m{-}3$
respectively, both modulo $8m{-}4$.  

For all the algebras we can summarize the commutation-relations within
$\ck_\bo$ in the surprisingly simple general expressions: 
\be
\begin{array}{rcr}
{}[\psi_i,\psi_j] & = & 2 \L^{i+j\over 2}  \\
{}[\bar{\psi}_i,\bar{\psi}_j] & = & - 2 \L^{i+j\over 2} \\
{}[\psi_i,\bar{\psi}_j] & = & {0~~~~}
\end{array}
\lab{comrel} 
\ee
The elements on the right-hand side of these equations form the
subset $\cz'\subset\cz$. 
In appendix \ref{sect:comrel} we find that in the case of 
$D(m{+}1,m)^{(1)}$ there is one series of elements in
$\cz\setminus\cz'$, which we denote by $\tilde{c}_{2m+4pm},\,p\in\fl{Z}$,
while in all other cases we have $\cz'=\cz$.  
We recall that in the algebra
$A(m,m)$ we have the equivalence relation $\one \sim 0$ where $\one$
is the $(m{+}1)\times(m{+}1)$ unit matrix~\footnote{We will use the
notation $\one$ for the unit matrix in any dimension}, which implies
that in $A(m,m)^{(1)}$ and $A(2m,2m)^{(4)}$ 
we have $\L^h=0$, where $h$ is the Coxeter
number. 
For the elements $\xi_i$ and $\bar{\xi}_i$ in $A(2m{-}1,2m{-}1)^{(2)}$,
we find 
\bea
& &[\xi_i,\xi_j] = 2 \L^{i+j\over
2}  \nn  
& &[\bar{\xi}_i,\bar{\xi}_j] = - 2 \L^{i+j\over 2} \nn
& &[\xi_i,\bar{\xi}_j] = {0~~~~} \nn
& &[\xi_i,\psi_j] = [\xi_i,\bar{\psi}_j]= [\bar{\xi}_i,\psi_j]
=[\bar{\xi}_i, \bar{\psi}_j]= 0
\lab{comrel2} 
\ena    

Combining these relations with theorem \ref{theo1}, we immediately have
explicit expressions for the algebra generated by the 
fermionic flows and the flows related to elements in $\cz$. 
Defining $D_{i} \equiv D_{\psi_i}$, $\bar{D}_i \equiv
D_{\bar{\psi}_i}$ and
$\partial_i = D_{\L^{i\over 2}}$: 
\be
\begin{array}{rcr}
{}[D_i,D_j] & = & 2 \partial_{i+j}  \\
{}[\bar{D}_i,\bar{D}_j] & = & - 2 \partial_{i+j} \\
{}[D_i,\bar{D}_j] & = & 0
\end{array}
\lab{comrel3} 
\ee
and similarly for the flows corresponding to eq.~\rf{comrel2}. 

Not all of these flows are supersymmetric, {\it i.e.\/}, not all 
are compatible with the supersymmetry transformation. 
According to theorem \ref{theo3}, the supersymmetric flows are
those corresponding to elements in $\Re = \ker(\ad\Psi)$,
{\it i.e.\/}, the elements in the centre plus
$\bar{\psi}_i,\,\xi_i,\, \bar{\xi}_i\in\ck_\bo$, with  corresponding
flows $\partial_i$, $D_{\tilde{c}_{2m+4pm}}$ in the case of
$D(m{+}1,m)^{(1)}$,
$\bar{D}_i$, $D_{\xi_i}$ and $D_{\bar{\xi}_i}$.
We have used the fact that for the principal hierarchies, the only
even elements which commute with $\Psi$ are those in the centre, or
$(\ck_\bz\setminus\cz)\cap\Re=\emptyset$, as is evident in appendix
\ref{App}. 

We have given here the commutation relations of the subalgebra 
$\ck_\bo\cup\cz$ of $\ck$. Using the explicit
expressions given in the appendices it is straightforward to find the
commutation relations for all elements of $\ck$, and thereby also 
the algebra generated by all the possible flows. 
The result is that $\cal K$ is a subalgebra of the
principal super Heisenberg algebra of $a_{\infty| \infty}$ without
central extension, which is isomorphic to the affine superalgebra
$\widetilde{gl}_{1|1}$~\cite{KacLeur,Fock}. This generalizes the well
known role of the principal Heisenberg algebra in the KP hierarchy and
in the original DS (bosonic) hierarchies. We expect that this
relationship helps to derive a tau-function formalism for these
hierarchies in analogy with refs.~\cite{Tau,Wilson,KW} and,
especially, to construct solutions for the equations of the hierarchy
using super vertex operator representations; in particular, multi-soliton
solutions.  

\subsection{Special Cases}

\subsubsection{$A(m,m)^{(1)}$} 

The hierarchies related to the algebras $A(m,m)^{(1)}$ have been
studied previously in e.g. \cite{DelGal,InKa2}, and it is known that in
these cases the supersymmetry can in fact be extended to an $N=2$
supersymmetry. 
In our formalism, there is a very natural explanation for this fact:
we see from the list above that there is a fermionic element 
$\bar{\psi}_1\in\ck$ which satisfies $\bar{\psi}_1^2 = -\L$. 
It follows from lemma \ref{lem4} that $\bar{D}_1$, 
the corresponding fermionic flow, is a {\em local} flow, and from
theorem \ref{theo1} 
that $(\bar{D}_1)^2 = -\partial_2 = -\partial_x$, and it is therefore
clear that 
$D_1=D$ and $i \bar{D}_1$ satisfies the $N=2$ supersymmetry algebra. 

\subsubsection{$A(2m{-}1,2m{-}1)^{(2)}$} 

In this case, as mentioned above, $\tilde{\ck}$ in eq.~\rf{Ktilde}
contains fermionic elements. According to eq.~\rf{ConsL},  this
implies that we have an infinite number of {\em local} conserved
fermionic quantities.  Moreover, explicit calculations show that
$[\ck_\bz,\xi_i]= [\ck_\bz,\bar{\xi}_i]= 0$. This, together with the commutations relations~\rf{comrel2}, shows that 
$[\xi_{2m{-}1},\ck^{<0}]\subset \ck^{<0}$, and therefore according to
lemma \ref{lem4} the corresponding flow $D_{\xi_{2m{-}1}}$ is local. 

\sect{Example: SKdV}
\label{SKdV} 

We will demonstrate the application of our construction in the example
of SKdV, the supersymmetric version of the usual KdV hierarchy. This
hierarchy has previously been studied by many authors, mainly in the
formalism of super pseudo-differential operators
\cite{ManRad,SKdV,DaMa,Ra,Mas}, but also using Lie superalgebras
\cite{InKa1}. There are several advantages in taking SKdV as example: it is one of the simplest possible
examples; we can compare the results of our construction with the results
obtained using other methods; and we can demonstrate to which extent our
method gives new results even for a hierarchy that has already been
considerably studied. 

The SKdV hierarchy is the supersymmetric hierarchy defined by taking
$\Psi$ to be the principal element in $A(0,1)^{(2)}$
\cite{InKa1}~\footnote{In \cite{InKa1}, the authors consider the
algebra $C(2)^{(2)} = osp(2,2)^{(2)}$. For simplicity we have chosen
to consider instead $A(0,1)^{(2)}$, since $C(2)\simeq A(0,1)$ and $A(0,1)$
is realized in terms of $3\times 3$ matrices, while we need $4\times 4$
matrices to realize $C(2)$.}, and to take $d_1$  to be the principal
grading $d_1 = (1,1)$, and $d_0$ to be the homogeneous grading,
$d_0=(1,0)$ 

Elements $x\in A(0,1)^{(2)}$ can be realized as $3\times 3$ matrices of
the form: 
$$
x = \left ( \begin{array}{c|cc}  
2 \l\, u  & \g_1 + \l\, \g_2 & \g_3 + \l\, \g_4
\\ \hline 
-\g_3+\l\,\g_4 & \l\, u + j_0 & j_+ \\
+\g_1 - \l\, \g_2 & j_- & \l\, u - j_0 
\end{array} \right ) 
$$ 
where $\g_i,\,u$ and $j_i$ are polynomials in $\l^2$ and $\l^{-2}$. 
We take $\Psi$ and $\L$ to be respectively: 
$$
\Psi = \left ( \begin{array}{c|cc} 
0 & \l & -1 \\ \hline 
1 & 0 & 0 \\ 
-\l & 0 & 0 \end{array} \right ) \hspace{20mm} 
\L = \left ( \begin{array}{c|rr} 
2\l  & 0 & 0 \\ \hline 
0 & \l  & -1 \\ 
0 & -\l^2 & \l \end{array} \right ) = \Psi^2
$$
In this realization 
we can write the principal gradation $d_1$ as 
ad($e_{2,2} - e_{3,3} + 2 \l \partder{}{\l}$), where $e_{i,j}$ denotes the
matrix with 1 in the $i,j$'th position, and zero elsewhere, while the
homogeneous gradation $d_0$ is ad($\l \partder{}{\l}$). 

$A(0,1)^{(2)}$ is one of the two only examples where $\L$ is not
the direct sum of two terms which are each a principal element in a simple
part of the even subalgebra (the other is $D(2,1)^{(1)}$). Instead $\L$ is
here a direct sum of a part which is principal in $A_1^{(1)}$, and a part
which lies in the $U(1)^{(1)}$ subalgebra. This gives only trivial
modifications to the procedure, however. 

In this case, $\ck_\bz$ is generated by the elements
$$
c_{2+4i} = \l^{2i} \left ( \begin{array}{c|cc} 
2\l & 0 & 0 \\ \hline 
0 & \l & 0 \\
0 & 0 & \l \end{array} \right ) \hspace{20mm} 
d_{2+4i} = \l^{2i} \left ( \begin{array}{c|cc} 
0 & 0 & 0 \\ \hline 
0 & 0 & 1 \\
0 & \l^2 & 0 \end{array} \right )
$$ 
where $\L=c_2-d_2$, while $\ck_\bo$ is generated by 
$$
\psi_{1+4i} =  \l^{2i} \left ( \begin{array}{c|cc} 
0 & \l & -1 \\ \hline 
1 & 0 & 0 \\
-\l & 0 & 0 \end{array} \right ) \hspace{20mm} 
\bar{\psi}_{3+4i} = \l^{2i} \left ( \begin{array}{c|cc} 
0 & -\l^2 & \l \\ \hline 
\l & 0 & 0 \\
-\l^2 & 0 & 0 \end{array} \right )
$$
and $\cz\subset \ck_\bz$ is generated by the elements $b_{2+4i}=\l^{2i}
\L$. They satisfy
\bea
& &[\psi_{1+4i}, \psi_{1+4j}] = 2 b_{2 + 4(i+j)}, \qquad
[\bar{\psi}_{3+4i}, \bar{\psi}_{3+4j}] = -2 b_{2 + 4(i+j+1)} \nn
\noalign{\vskip 0.2truecm}
& & [d_{2+4i}, \psi_{1+4j}] = -
\bar{\psi}_{3+4(i+j)},
\qquad [d_{2+4i}, \bar{\psi}_{3+4j}] = - {\psi}_{1+4(i+j+1)}, \nn
\ena
and it is worth noticing that the conventions we are using in this
particular case are different from the general ones in the previous 
section (for example see eq.~\rf{comrel} and notice that $b_{2+4i}\not=
\Lambda^{1+2i}$).

We choose the Drinfel'd-Sokolov gauge to be of the form 
$$
q_V = \left ( \begin{array}{c|cc} 
0 & 0 & 0 \\ \hline 
0 & 0 & 0 \\
0 & \Phi(\tilde{x}) & 0 \end{array} \right ) 
$$
$\Phi(\tilde{x})$ is a superfield which we can expand on fields as 
$\Phi(\tilde{x})=\xi(x)+\th u(x)$, where $u(x)$ is the usual KdV
function and $\xi(x)$ is its superpartner. 

Using the definition~\rf{SusyFlow} for the bosonic flows
corresponding to the elements of $\cz$, we find the flow corresponding to 
$b_6$ to be: 
$$
\partial_6 \Phi(\tilde{x}) = -D^6 \Phi + 3 D^2(\Phi D\Phi)
$$
which is the SKdV equation, and we find $H$, defined in eq.
\rf{Abelianize}, to be 
$$
H = \hf \Phi d_{-2} - {1\over 8} \Phi D\Phi d_{-6} + \cdots
$$ 
where the ellipsis symbolises terms with lower grades, so 
the local conserved currents related to the first two bosonic flows are
$\Phi$ and $\Phi D\Phi$, respectively. 

In order to calculate the first non-trivial
supersymmetric fermionic flow, $\bar{D}_3$, we need to find the non-local
quantity $\omega$, defined in eq.~\rf{AbelPlus}. For the first few grades,
we find:
\bea
\omega &=& -\hf D^{-2} \Phi \bar{\psi}_{-1} + 
{1\over 4} D^{-2} (\Phi D^{-2}\Phi) b_{-2} + 
\hf D^{-1} \Phi d_{-2} - 
{1\over 8} D^{-1} ( \Phi D^{-2}\Phi) \psi_{-3} + \nn
&& \hspace{-5mm} + {1\over 96} D^{-1} \left 
( 12 D^{-1}(\Phi D\Phi) - 3 \Phi D^{-1}(\Phi D^{-2}\Phi) + 
\Phi D^{-1}\Phi D^{-2}\Phi \right )\bar{\psi}_{-5} + \cdots 
\lab{omega}
\ena
where again the ellipsis denotes terms of lower grade. 
Using this and eq.~\rf{SusyFlow}, we 
find the first non-trivial fermionic flow to be: 
$$ 
\bar{D}_3 \Phi = - D^3 \Phi + 2D^2 \Phi D^{-2} \Phi 
- D\Phi D^{-1}\Phi 
$$
This is exactly the non-local flow found in \cite{DaMa}, where the
fermionic flows were defined using pseudo-differential operators as flows
related to the fourth root of the Lax operator. We can verify by
direct calculation that this flow satisfies 
$\bar{D}_3^2= - \partial_6$. 

In section~\ref{sec4} we have shown that if we impose suitable
boundary conditions on the fields, in this case assuming that
$\Phi(\tilde{x})$  goes sufficiently fast to zero for $x \rightarrow \pm 
\pm\infty$, then  the components of $\omega|_{x=\pm \infty}$ are conserved
quantities under all bosonic flows. It is instructive to check explicitly
that this is true with respect to $\partial_6$, since in order to show
conservation we already need to use partial integration with non-local
quantities, as described on page~\pageref{eq:partial-int}. We have done
this calculation down to grade $-5$. In this case the conservation depends
on a considerable number of  cancellations of terms, giving a non-trivial
check of the  conservation of $\omega|_{x=\pm \infty}$. Here we will
restrict ourselves to show the conservation of the components of
grade~$-1$, $-2$ and~$-3$ only. 

According to~\rf{Jchi}, the conserved quantity associated with
$\Psi=\Psi_1$ is
\be
J_{\Psi} = \int_{-\infty}^{+\infty} dx\> 
\partial_x\>  {\rm str\/}\> \bigl(\Psi {\rm
e\/}^{\omega}\bigr) = \int_{-\infty}^{+\infty} dx\> 
\partial_x\>  {\rm str\/}\> \bigl(\chi \omega\bigr) =  -2
\int_{-\infty}^{+\infty} dx\> \Phi
\ee
which, using~\rf{Split}, gives rise to two scalar conserved quantities
\be
I_{\Psi} = -2 \int d\tilde x\> D^{-1} \Phi, \qquad
I_{[\Psi,\Psi]}= 2I_\Lambda = 2 \int d\tilde x\> \Phi
\ee
Using proposition~\ref{Prop5}, $I_\Lambda$ is a supersymmetric
Grassmann-even conserved quantity while $I_{\Psi}$ is a
non-supersymmetric fermionic conserved quantity. This agrees with the
results of~\cite{DaMa}, where $I_\Psi \propto J_{1/2}$ and $I_\Lambda
\propto H_1$, which is the Hamiltonian corresponding to the even flow
$\partial_2 = \partial_x$.

$\omega$ has two different terms of principal degree~2, which gives rise
to two independent Grassman-even conserved quantities. The first one is
$I_\Lambda$, and the second is given by
\be
J_{c_2} = \int_{-\infty}^{+\infty} dx\> 
\partial_x\> {\rm str\/}\> \bigl(c_2 {\rm
e\/}^{\omega}\bigr) = \int_{-\infty}^{+\infty} dx\> 
\partial_x\>  {\rm str\/}\> \bigl(c_2 \omega\bigr) =  {1\over2}
\int_{-\infty}^{+\infty} dx\> \Phi D^{-2}\Phi
\ee
which has two scalar components. It is straightforward to check that
\be
J_{c_2} = I_{c_2} + \theta I_{\bar{\Psi}_3}
\ee
which agrees with eq.~\rf{Split} since $[c_2,\Psi]= -\bar{\Psi}_3$.
Then,
\be
I_{c_2} = {1\over2} \int d\tilde x\> D^{-1} \bigl(\Phi D^{-2}\Phi\bigl)
\ee
provides a non-supersymmetric ($[c_2,\Psi]\not=0$) Grassmann-even
conserved quantity. Actually, this non-local quantity was not
mentioned in~\cite{DaMa} and has not previously been shown to be a
conserved quantity. Our construction ensures that it is conserved by
all the bosonic flow equations of the hierarchy. In particular, 
\bea 
\partial_6 J_{c_2}  
& = &  {1\over2} \int_{-\infty}^{+\infty} dx~ \partial_6 (\Phi D^{-2}\Phi )
\nn & = & {1\over2}\int_{-\infty}^{+\infty} dx~ 
\Bigl[ (D^6 \Phi - 3 D^2(\Phi D\Phi))D^{-2}\Phi + 
\Phi (D^4 \Phi - 3 (\Phi D\Phi))\Bigr] \nn
& = & {1\over2} \int_{-\infty}^{+\infty} dx~  2 \Phi D^4 \Phi  =\> 0
\nonumber
\ena 
where we have used partial integration, as well as the fact that
$\Phi\Phi=D^2\Phi D^2\Phi=0$. This ensures the conservation of both
$I_{c_2}$ and $I_{\bar{\Psi}_3}$. 

Since $[\bar{\Psi}_3, \Psi]=0$, $I_{\bar{\Psi}_3}$ is a
non-local supersymmetric fermionic conserved quantity:
\be
I_{\bar{\Psi}_3} = J_{\bar{\Psi}_3} = \int d\tilde x\> \bigl(\Phi
D^{-2}\Phi\bigl)
\ee
As explained in page~\pageref{scale}, this conserved quantity has scale
dimension $3/2$ and, hence, should be related to $J_{3/2}$ in~\cite{DaMa}.
The precise connection is 
\be
J_{3\over2} = - I_{\bar{\Psi}_3} - 4\int d\tilde x\> D\bigl(\omega(-1)
\omega_d(-2)\bigr)
\ee
where $\omega(-1)$ and $\omega_d(-2)$ are the coefficients of
$\bar{\Psi}_{-1}$ and $d_{-2}$ in eq.~\rf{omega}, respectively. The
conservation of $J_{3\over2}$ follows directly from the conservation of
the components of $\omega|_{x=\pm\infty}$.


\sect{Conclusions}
\label{sec7}

For the entire class of supersymmetric Drinfel'd-Sokolov integrable
hierarchies constructed by Delduc and Gallot in~\cite{DelGal}, we have
defined an infinite sequence of additional Grassmann-odd and
Grassmann-even flow equations, which are given by non-local equations,
{\em i.e.\/}, which are not defined by super-differential polynomials of
the dynamical variables. In this sense, they have to be seen as
generalizations of the fermionic flows constructed by Dargis and Mathieu
for the supersymmetric KdV equation in~\cite{DaMa}. All these flows
commute with the bosonic ones originally considered in~\cite{DelGal} and,
hence, provide new fermionic and bosonic symmetry transformations for the
hierarchy. In particular, one of the fermionic flows corresponds
precisely to the built-in supersymmetry transformation and, in some
cases, others provide additional supersymmetries. The new flow equations
do not commute among themselves; instead, they close an infinite
non-abelian superalgebra. 

The supersymmetric hierarchies of~\cite{DelGal} are associated with a
(twisted) loop superalgebra $\cal A$, and two compatible
gradations~$d_1$ and~$d_0$. They are defined by means of 
fermionic Lax operators of the form ${\cal L} = D + q(x,\theta) +\Psi$
where $q(x,\theta)$ is a $N=1$ superfield taking values in a subspace of
$\cal A$, $D$ is the superderivative, and $\Psi$ is a constant
$d_1$-graded element of $\cal A$ whose square $\Lambda = [\Psi, \Psi]/2$
is semi-simple. Then, following our construction, there is a
non-local flow associated with each element in ${\cal K}= {\rm
Ker\/}({\rm ad\/} \Lambda)$ with non-negative $d_1$-grade, {\em i.e.\/},
in ${\cal K}^{\geq0}$, and these flows close an infinite superalgebra
isomorphic to ${\cal K}^{\geq0}$. In contrast, the mutually
commuting flows originally considered by the authors of~\cite{DelGal} are
associated with the elements in the centre of ${\cal K}^{\geq0}$, which
contains only even elements. Since the built-in supersymmetry
transformation
$\delta^{\rm SUSY}_\eta = \eta D$ is identified with the flow equation
associated with $\Psi$, it is easy to characterise the flows which are
supersymmetric: those associated with the elements in ${\cal
K}^{\geq0}$ that commute with $\Psi$. Of course, all the commuting
bosonic flows of~\cite{DelGal} are supersymmetric.

We have also constructed an infinite series of non-local quantities which
are conserved by all the commuting bosonic flows associated with the
centre of ${\cal K}^{\geq0}$. To be precise, there is a non-local
conserved quantity for each element in ${\cal K}^{\geq0}$, and only those
conserved quantities associated with the elements that commute with
$\Psi$ are supersymmetric. Therefore, there is a one-to-one relationship
between non-local flow equations and non-local conserved quantities,
which suggest that the latter could be understood as the generators of
the former. In the case of the supersymmetric KdV equation, this
relationship was established in the work of Dargis and
Mathieu~\cite{DaMa} and Ramos~\cite{Ra} by using the super
pseudo-differential Lax operator formalism. In our case, the precise
nature of this relationship is one of the open questions that is left to
be the subject of future work.

As an example of our construction, we have worked out in detail the
structure of the non-local flow equations and conserved currents for the,
so called, principal hierarchies, which are associated with the affine
superalgebras that have a fermionic simple root system; namely, those
listed in table~\ref{FSRS}. Then, $d_1$ is the principal gradation and
$\Psi$ is the cyclic element, {\em i.e.\/}, the sum of the (affine)
simple root generators. Actually, these hierarchies can be seen as the
direct generalization of the original Drinfel'd-Sokolov construction,
and their commuting bosonic flows were defined first by Inami and
Kanno~\cite{InKa1,InKa2}. In order to proceed, we have constructed
explicit matrix-representations for the  affine Lie superalgebras with
fermionic simple root systems which, as far as we know, are not
available in the literature. Using them, we have shown that ${\cal K}=
{\rm Ker\/}({\rm ad\/} \Lambda)$ is a subalgebra of the infinite
superoscillator algebra constructed by Kac and van de Leur
in~\cite{KacLeur}, {\em i.e.\/}, of the principal Heisenberg
superalgebra of $a_{\infty| \infty}$ without central
extension~\cite{Fock}. The precise form of the subalgebra depends on the
particular superalgebra we have started with. This generalises the well
known role of the oscillator and Heisenberg algebras in the KP
hierarchy~\cite{KP} and the generalized hierarchies of KdV
type~\cite{miramontesetal,Tau,KW}. It also suggests the possibility of
understanding our hierarchies of local and non-local flows as reductions
of the super KP hierarchy of super Hirota bilinear equations constructed
in the second article of~\cite{KacLeur}.

Finally, let us comment about the construction of solutions for the
equations of these hierarchies. According to~\rf{Laxes}, a particular
solution is specified by $\Theta$, which satisfies the system of first
order super differential equations given by~\rf{Flows}. In order to
solve it, consider the decomposition ${_+{\cal A}} = {_+{\cal A}}^{<0}
\oplus {_+{\cal A}}^{\geq0}$, where ${_+{\cal A}}^{<0}$ (${_+{\cal
A}}^{\geq0}$) is the subalgebra formed by the elements with negative
(non-negative) $d_1$-grade. Next, let $G_-$ and $G_+$ be the subgroups
obtained by (formally) exponentiating the subalgebras ${_+{\cal
A}}^{<0}$ and ${_+{\cal A}}^{\geq0}$, respectively, and define $G =G_-\>
G_+$, which is the super analogue of the big cell of a loop
group~\cite{Wilson}; if $g$ belongs to $G$, we shall write $g= g_- g_+$
for its unique factorization with $g_\pm \in G_\pm$. Then, the general
solution of~\rf{Flows} is
\be
\Theta = \bigl(\Gamma_0 g\bigr)_- \>,
\lab{Solution}
\ee
where $g$ is any constant element in $G_-$ and $\Gamma_0\in G_+$ is the
solution of the associated linear problem corresponding to $\Theta=\one$:
\be
{\cal L}_u \Big|_{\Theta=\one} \Gamma_0 = \bigl(D_u - \hat{u}\bigr)
\Gamma_0 =0\>;
\lab{Vacuum}
\ee
in the parlance of the second and third articles in~\cite{Tau}, $\Gamma_0$
is the `vacuum' solution of the associated linear problem.
Eq.~\rf{Solution} provides an infinite set of solutions parametrized by
the elements of $G_-$. By analogy with the bosonic hierarchies, the
elements obtained by exponentiating super vertex operators should be
related to the multi-soliton solutions~\cite{Tau}.

Eq.~\rf{Vacuum} can be easily solved for the principal hierarchies. To
be specific, let us restrict ourselves to the flows associated with the
elements of the form $\psi_{-1+2i}$, $\bar{\psi}_{1+2i}$, and $b_i=
\Lambda^{i\over2}$, which is a subset of ${\cal K}_{\bar1}\cup {\cal
Z}\subset {\cal K}^{\geq0}$; the resulting expressions can be easily
modified to cover the complete set of flows. In order to
solve~\rf{Vacuum}, it is convenient to introduce and infinite number of
even times $t_i$ associated with $b_i$, and two infinite sets of odd
times $\theta_i$ and $\tau_i$ associated with $\psi_i$ and 
$\bar{\psi}_i$, respectively. Then, since $D_1= D$, the flows can be
realized on the set of functions of these variables according to
\bea
& & \partial_i = {\partial\over \partial t_i}, \qquad D_1 = {\partial
\over \partial \theta_1} + \theta_1 {\partial\over \partial t_2} \nn 
& & D_i = {\partial \over \partial \theta_i} + 2\theta_1
{\partial\over
\partial t_{i+1}} + \sum_{j\not=1} \theta_j {\partial\over
\partial t_{i+j}}, \quad i\not=1 \nn
& & \bar{D}_i = {\partial \over \partial \tau_i} - \sum_{j} \tau_j
{\partial\over \partial t_{i+j}}\>,
\ena
which trivially satisfy eq.~\rf{comrel3}. Using these variables, the
vacuum solution is
\be
\Gamma_0 = \exp \sum_i \bigl( t_i b_i - \theta_i \psi_i - \tau_i
\bar{\psi}_i \bigr) \>,
\ee
where $\theta_1=\theta$ and $t_2 =x$.

\newpage

\noindent\centerline{\large\bf Acknowledgments} 

\vspace{0.5truecm}
We would like to thank J.~Evans for valuable discussions and
F.~Delduc for kindly sending us an early version of
ref.~\cite{DelGal}. This research is supported partially by the EC 
Commission via the TMR Grant FMRX-CT96-0012, CICYT
(AEN96-1673), and DGICYT (PB96-0960).

\vspace{2 truecm}

\appendix
\sect{Proof of Proposition~1}
\label{AppA}

~\indent
First of all, since ${\rm Ker\/}({\rm
ad\/}
\Psi)\subset {\cal K}$, let us choose a vector subspace $\cal V$
such that
\be
{\cal K} = {\rm Ker\/}({\rm ad\/} \Psi) \oplus {\cal V}\>,
\ee
which implies that 
\be
\Im = {\rm Im\/}({\rm ad\/} \Psi) \cap {\cal K} = [\Psi,
{\cal V}]\>.
\ee
Moreover, if $\Re = {\rm Ker\/}({\rm ad\/} \Psi)$, it is
straightforward to check that
\be
\Im \subseteq \Re\>, \qquad [\Re, \Re] \subset \Re\>, \qquad
[\Im, \Im]\subset \Im\>.
\lab{Alg}
\ee

The proof of the Proposition goes by induction on the $d_1$-grade
of the components of~\rf{AbelPlus}. The highest grade component
reads
\be
H^{k-1} + [\hat\omega^{-1}, \Psi] =0\>,
\lab{ConstA}
\ee
Which provides the first constraint, $H^{k-1} \in \Im$, and
uniquely fixes the component of $\omega^{-1}$ in $\cal V$, namely
$\omega^{-1}_{\cal V}$, as a local function of $H^{k-1}$. Then,
consider the transformation
\be
{\rm e\/}^{\hat \omega^{-1}_{\cal V}} (D + \Psi + H) {\rm e\/}^{-
\omega^{-1}_{\cal V}} = D +\Psi + H_{(1)}\>;
\ee
it is obvious that $H_{(1)}^{k-1}=0$. Next, suppose that $H^{>j}
=0$ for $j\geq0$ and consider the transformation
\be
{\rm e\/}^{\hat \omega^{-k+j}_{\cal V}} (D + \Psi + H) {\rm
e\/}^{- \omega^{-k+j}_{\cal V}} = D +\Psi + H^j + [\hat
\omega^{-k+j}_{\cal V}, \Psi] + \cdots \>,
\ee
where the ellipsis indicates terms with $d_1$-grade $<j$. Therefore,
if $H^{j}$ is constrained such that $H^{j} \in \Im$, the condition
that the component of grade $j$ in the right-hand side vanishes
uniquely fixes $\omega^{-k+j}_{\cal V}$.

Following this procedure, we can construct a transformation 
\be
{\rm e\/}^{\omega^\ast} = {\rm e\/}^{\omega_{\cal
V}^{-k}}{\rm e\/}^{\omega_{\cal V}^{-k+1}} \cdots {\rm
e\/}^{\omega_{\cal V}^{-1}}
\ee
such that
\be
{\rm e\/}^{\hat \omega^\ast} (D + \Psi + H) {\rm
e\/}^{- \omega^\ast} = D +\Psi + H^\ast
\lab{ConstB}
\ee
and ${H^\ast}^{\geq0}=0$. 

Next, let us constrain the components of $H^\ast$ in $\cal V$ with
$d_1$-grade $\geq -k$ to vanish, {\it i.e.\/},
\be
{H^\ast_{\cal V}}^{\geq-k}=0
\qquad {\rm or} \qquad {H^\ast}^{\geq-k} \in \Re\>,
\lab{ConstC}
\ee
and consider the transformation
\bea
& &{\rm e\/}^{\hat \omega_{\cal V}^{-k-1}} {\rm e\/}^{\hat
\omega_{\Re}^{-1}} (D + \Psi + H^\ast) {\rm e\/}^{-
\omega_{\Re}^{-1}} {\rm e\/}^{- \omega_{\cal V}^{-k-1}}
\nn
& &\qquad\qquad= {\rm e\/}^{\hat
\omega_{\Re}^{-1}} \bigl(D +H^\ast\bigr) {\rm e\/}^{-
\omega_{\Re}^{-1}} +\Psi - D\omega_{\cal V}^{-k-1} +
[\hat \omega_{\cal V}^{-k-1}, \Psi] + \cdots\>,
\ena
where the ellipsis indicates terms with $d_1$-grade $< -k-1$. Then,
taking into account~\rf{Alg}, the condition that the terms with
$d_1$-grade $-1$ and the terms in $\cal V$ with grade $-k-1$ vanish
require
\bea
& & {H^\ast_{\cal V}}^{-k-1} - D\omega_{\cal V}^{-k-1} =0 \nn
& & {H^\ast}^{-1} - D\omega_{\Re}^{-1} + [\hat\omega_{\cal
V}^{-k-1}, \Psi] =0\>,
\ena
which uniquely fixes $\omega_{\cal V}^{-k-1}$ and $\omega_{ \Re}^{-1}$
as non-local functions of ${H^\ast_{\cal V}}^{-k-1}$ and ${H^\ast
}^{-1}$. Finally, suppose that ${H^\ast}^{>-j} =  {H^\ast_{\cal
V}}^{>-k-j}=0$ for some $j>1$, and consider
\bea
& &{\rm e\/}^{\hat \omega_{\cal V}^{-k-j}} {\rm e\/}^{\hat
\omega_{\Re}^{-j}} (D + \Psi + H^\ast) {\rm e\/}^{-
\omega_{\Re}^{-j}} {\rm e\/}^{- \omega_{\cal V}^{-k-j}}
\nn
&& \qquad\qquad= {\rm e\/}^{\hat
\omega_{\Re}^{-j}} \bigl(D +H^\ast\bigr) {\rm e\/}^{-
\omega_{\Re}^{-j}} +\Psi +  - D\omega_{\cal V}^{-k-j} +
[\hat \omega_{\cal V}^{-k-j}, \Psi] + \cdots\>.
\ena
In this case, the vanishing of the terms with
$d_1$-grade $-j$ and the terms in $\cal V$ with grade $-k-j$
is equivalent to
\bea
& &{H^\ast_{\cal V}}^{-k-j} - D\omega_{\cal V}^{-k-j} =0 \nn
& &{H^\ast}^{-j} - D\omega_{\Re}^{-j} + [\hat\omega_{\cal
V}^{-k-j}, \Psi] =0\>,
\ena
which uniquely fixes $\omega_{\cal V}^{-k-j}$ and
$\omega_{\Re}^{-j}$.

To summarize, we have constructed a transformation
\be
{\rm e\/}^{-\omega} = {\rm e\/}^{-\omega_{\cal
V}^{-1}} \cdots {\rm e\/}^{-\omega_{\cal V}^{-k}}
{\rm e\/}^{-\omega_{\Re}^{-1}}{\rm e\/}^{-\omega_{\cal V}^{-k-1}}
{\rm e\/}^{-\omega_{\Re}^{-2}} {\rm e\/}^{-\omega_{\cal V}^{-k-2}}
\cdots
\lab{Fin}
\ee
that satisfies~\rf{AbelPlus}, {\em i.e.\/},
\be 
{\rm e\/}^{\hat \omega}\bigl( D + \Psi + H\bigr) {\rm
e\/}^{-\omega} = D + \Psi\>,
\ee 
which completes the proof of Proposition~1; 

\sect{Lie superalgebras}

In this appendix we will recall a few basic facts about Lie
superalgebras which are of special relevance to this paper. 
For a complete introduction to the subject we refer to
e.g. \cite{Liesuper,Leur}  

\subsection{Bilinear Form}
\label{AppB}

A Lie superalgebra is a $\fl{Z}_2$-graded algebra 
$\cg=\cg_\bz\oplus\cg_\bo$, equipped with a graded commutator, 
which satisfies the graded Jacobi identity. 
We are here interested only in the {\em contragredient} or {\em basic
classical} Lie superalgebras, which are those that are equipped with a
non-degenerate bilinear form $\langle\cdot,\cdot\rangle$ which is
invariant, {\it i.e.\/}: 
$$
\langle X,[Y,Z]\rangle = \langle[X,Y],Z\rangle
$$
Consider such a Lie superalgebra $\cg$. 
Given a matrix-representation $R$ of $\cg$, the bilinear form
in this representation is defined by 
$$
\langle X,Y\rangle_R = \str(R(X)R(Y))
$$ 
where str is the supertrace. A standard matrix representation can be
written in the form: 
$$ 
X = \left ( \begin{array}{cc} a & \a \\ \b & b \end{array} \right ) 
$$ 
where $a,\,b,\,\a$ and $\b$ are matrices, 
and the even subspace is formed of matrices of the form 
$\left ( \begin{array}{cc} a & 0 \\ 0 & b \end{array} \right )$; in
this representation the supertrace is defined by 
$$
\str(X) = \tr(a) - \tr(b)
$$

All bilinear forms are equal up to normalization. The Killing form is
the bilinear form in the adjoint representation; it may be identically
zero. 

Note that while the bilinear form is defined only on the superalgebra,
it is always possible to work in some matrix representation, and in
that case the supertrace is defined on the universal enveloping
algebra. 

If $h$ is an element of the Cartan subalgebra $H\subset \cg$ and
$E_\a$ is a root generator, then we have 
$$
[h,E_\a] = \a(h) E_\a
$$ 
where $\a(\cdot)$ is a linear functional on $H$. We denote by $h_\a$
the element in $H$ such that $\a(h) = \langle h,h_\a\rangle$. We have
$$
[E_\a,E_{-\a}] = \langle E_\a,E_{-\a}\rangle h_\a,
$$
where $\langle E_\a,E_{-\a}\rangle\not = 0$. 

The bilinear form gives rise to an `inner product' (which is in
general not positive definite) by 
$$
\a\cdot\b = \langle h_\a,h_\b\rangle
$$
It follows directly that 
$$
[h_\a,E_\b] = \b(h_\a) \,E_\b = (\a\cdot\b) \,E_\b
$$

\subsection{Root Systems of Affine Lie Superalgebras.} 

In this section, we will recall some useful facts about the
root systems of 
Lie superalgebras. We will give the standard realizations of the root
systems of some finite Lie superalgebras, which is useful for the
calculation of the signature of the root-spaces, see page~\pageref{Dim},
and we will give the fermionic simple root systems, as well as 
explicit matrix realizations of the principal elements $\Psi$. 

Most of the construction in section~\ref{sec-5} can be done without
knowing the precise form of the principal element $\Psi$. In general, it
is sufficient to know that we can find a matrix representation where 
$\Psi$ takes the form 
\be
\Psi = \left ( \begin{array}{cc} 0 & \psi \\ \tilde{\psi} & 0
\end{array} \right ) 
\lab{psi}
\ee
where $\psi$ and $\tilde{\psi}$ are unspecified matrices, and that 
$$
\Psi^2 = \left ( \begin{array}{cc} \L_1 & 0 \\ 0 & \L_2 \end{array}
\right ) 
$$
where $\L_1$ and $\L_2$ are principal elements in the Lie algebras
$\cg_1$ and $\cg_2$, together with the knowledge that we have of the
structure of the bosonic case. There are, however, a few cases where
it turns out to be necessary to actually find an expression for 
$\Psi$ in order to do explicit calculations. Explicit calculations are
necessary to show that the `spurious' element of $\ck_\bz$ in the
case of $D(m+1,m)^{(1)}$ is an element of $\cz$, see page
\pageref{Dcentre}, and to find the extra elements of $\ck_\bo$ and
their commutation relations in the case of $A(2m-1,2m-1)^{(2)}$, see
page \pageref{Akernel}. 

At the same time, we are not aware of any reference in which one can
find explicit expressions for the matrix-representations of fermionic 
simple root systems of affine Lie superalgebras. 
Since Lie superalgebras with fermionic simple root systems are 
important in several applications of Lie superalgebras, for example in
supersymmetric Toda theory or supersymmetric hamiltonian reduction, 
we have chosen to give these expressions in some detail.

Note that there are other matrix realizations of Lie superalgebras
which may be useful when studying supersymmetric integrable
hierarchies, see \cite{De-etal}.   

In all cases 
$\ve_1,\ldots,\ve_n$ denotes an orthonormal basis of a
vector space $V_+$, while $\d_1,\ldots,\d_m$ denotes a basis of a
vector space $V_-$ with negative definite inner product, normalized so  
$\d_i\cdot\d_j=-\d_{ij}$ (where $\d_{ij}$ is the usual Kronecker
delta-symbol).
In addition, $e_{i,j}$ will
denote a matrix (the dimensions will be clear from the context) with
one in the $i,j$'th position, and zero elsewhere. 

\subsubsection{Root systems}
\paragraph{$A(n{-}1,m{-}1)$}~\\
even roots: $\ve_i-\ve_j,\,\d_i-\d_j,\,i\not=j$.  \\
odd roots: $\pm(\ve_i-\d_j)$.
\paragraph{$B(n,m)$}~\\
even roots: $\pm\ve_i\pm\ve_j,\,\pm\ve_i,\, 
\pm\d_i\pm\d_j,\,\pm 2\d_i,\,i\not=j$.  \\
odd roots: $\pm\ve_i\pm\d_j,\,\pm\d_i$.
\paragraph{$D(n,m)$} ~\\ 
even roots: $\pm\ve_i\pm\ve_j,\, 
\pm\d_i\pm\d_j,\,\pm 2\d_i,\,i\not=j$.  \\
odd roots: $\pm\ve_i\pm\d_j$.\\ 

For $B(n,m)$ and $D(n,m)$ the root space is $V_+\oplus V_-$. 
In the case of $A(n{-}1,m{-}1)$, the root space is the
subspace of $V_+\oplus V_-$ of co-dimension 1, orthogonal to 
$\sum_i \ve_i - \sum_j \d_j$. 

The upper limit on the dimension of $\phi(\ck_\bo)$ is given by 
2 min$(d'_+,d'_-)$, where $(d'_+,d'_-)$ is the signature of the root
space of the superalgebra. 
If we denote by $(d_+,d_-)=(\mbox{dim}(V_+),\mbox{dim}(V_-))$ the
signature of $V_+\oplus V_-$, then clearly 
min$(d'_+,d'_-)\le$min$(d_+,d_-)$, and we can therefore safely use the 
signature of $V_+\oplus V_-$ as the upper limit in the general case.
\label{VV}

We will now give the fermionic simple root systems and the explicit
matrices corresponding to the principal elements considered in this
paper. Let us remind the reader that the simple root system for an
untwisted affine Lie superalgebra can be chosen to consist of the
simple root system of the corresponding finite Lie superalgebra, with
the addition of an `affine root', which can be realized as 
$\l E_{-\th}$, where $\l$ is the loop-parameter, and $E_{-\th}$ is the
generator corresponding to minus the highest root. 
In the case of an affine Lie superalgebra twisted with an automorphism
of order $k$, the automorphism gives rise to a grading of the Lie
superalgebra $\ch = \sum_{i=0}^{k{-}1} \cg_i$ where $\cg_0$ is
invariant under $\t$, and each $\cg_i$ forms a
representation of $\cg_0$. As a simple root system of the twisted
affine Lie superalgebra we can take a simple root system of $\cg_0$ in
addition to the `affine root' which we can realize as 
$\l E_{-\th}$, where $E_{-\th}$ is in this case the (unique) lowest
weight in the $\cg_0$-representation provided by $\cg_1$. 

\subsubsection{$A(m,m)^{(1)}$}
Fermionic simple root system: 
$(\ve_i-\d_i),i=1,\ldots,m+1;\,(\d_i-e_{i+1}),i=1,\ldots,m$. Matrix
realization: 
\bea 
E_{\ve_i-\d_i} &=& e_{i,m+1+i}\nn
E_{\d_i-\ve_{i{+}1}} &=& e_{m+1+i,i+1}
\ena  
The realization of the `affine root' is 
$\l e_{2m{+}2,1}$. 
In this basis $\Psi$ takes the simple form: 
$$
\Psi = \left ( \begin{array}{cc} 
0 & \one \\ \L_{A_m} & 0 \end{array} \right ) 
$$
where $\L_{A_m}$ is the usual principal element in $A_m$:
$$
\L_{A_m} = \left ( \begin{array}{cccccc}
0 & 1 & 0 & \cdots & 0 & 0 \\
0 & 0 & 1 & \cdots & 0 & 0 \\
0 & 0 & 0 & \cdots & 0 & 0 \\
\vdots & \vdots & \vdots & \ddots & \vdots & \vdots \\
0 & 0 & 0 & \cdots & 0 & 1 \\
\l & 0 & 0 & \cdots & 0 & 0 \end{array} \right ) 
$$

\subsubsection{$A(2m{-}1,2m{-}1)^{(2)}$}
\label{Ax}
The invariant subalgebra is $D(m,m)=osp(2m,2m)$. The simple roots of
$D(m,m)$ are $(\ve_i-\d_{i+1}),\,i=1,\ldots,m-1;\, 
(\d_i-\ve_i),i=1,\ldots,m;\,(\d_m+\ve_m)$. The matrix realization of
these roots are: 
\bea 
E_{\ve_i-\d_{i+1}} & = & e_{4m{-}i,2m{+}1{-}i}+e_{i,2m{+}1{+}i} \nn
E_{\d_{i}-\ve_i} & = & e_{2m{+}1{-}i,4m{+}1{-}i}-e_{2m{+}i,i} \nn
E_{\d_{m}+\ve_m} & = & e_{m,3m{+}1}-e_{3m,m{+}1} 
\ena
while the matrix realization of the affine root is 
$\l(e_{2m,2m{+1}}-e_{4m,1})$, such that $\Psi$ takes the form of
eq. \rf{psi} with:
$$
\psi = \left (  
\begin{array}{clcc|ccc}
0 &      1 &        &  &&& \\
  & \ddots & \!\!\!\!\ddots &  &&& \\
  &        &        &1 &&& \\
  &        &        &0 &1&& \\ \hline
&&&&1&&\\
&&&&&\ddots&\\
\l&&&&&&1
\end{array} \right ) \hspace{20mm}
\tilde{\psi} = \left ( \begin{array}{ccc|rlcc}
\sm 1&&&&&&\\
&\ddots&&&&&\\
&&\sm 1&\sm 1&&&\\ \hline
&&&0&1&& \\
&&&&\ddots&\!\!\!\!\ddots& \\
&&&&&&1 \\
\sm \l&&&&&&0 \end{array} \right ) 
$$ 
where the blocks are $m\times m$ matrices. 

\subsubsection{$A(2m{-}2,2m{-}1)^{(2)}$}

The invariant subalgebra is $B(m{-}1,m)=osp(2m{-}1,2m)$. 
The simple roots of $B(m{-}1,m)$ are 
$(\ve_i-\d_{i{+}1}),i=1,\ldots,m{-}1;\,(\d_i-\ve_i),i=1,\ldots,m{-}1;
\,\d_m$.
The matrix realization is:
\bea 
E_{\ve_i-\d_{i{+}1}} & = & e_{4m{-}1{-}i,2m{-}i} + e_{i,2m{+}i} \nn
E_{\d_i-\ve_i} & = & e_{2m{-}i,4m{-}i} - e_{2m{-}1{+}i,i} \nn 
E_{\d_m} & = & e_{m,3m} - e_{3m{-}1,m} 
\ena 
while the realization of the affine root is 
$\l(e_{2m{-}1,2m}-e_{4m{-}1,1})$, 
such that $\Psi$ takes the form in eq. \rf{psi} with: 
$$
\psi = \left ( \begin{array}{clcc|clrc}
0&1&&&&&& \\
&\ddots&\!\!\!\!\ddots&&&&& \\
&&0&1&&&&\\ \hline 
&&&&1&&&\\ \hline 
&&&&0&1&&\\
&&&&&\ddots&\!\!\!\!\ddots&\\
\l&&&&&&0&1\end{array}\right )
\hspace{20mm} 
\tilde{\psi} = \left ( \begin{array}{rrr|c|ccc}
\sm 1&&&&&& \\
0&\ddots&&&&& \\
&\ddots&\sm 1&&&& \\
&&0&\sm 1&&& \\ \hline 
&&&&1&& \\
&&&&0&\ddots&\\
&&&&&\ddots&1\\
\sm\l&&&&&&0\end{array}\right )
$$
where the `big' blocks are $(m{-}1)\times m$-matrices for $\psi$,
and $m\times (m{-}1)$-matrices for $\tilde{\psi}$. 

\subsubsection{$B(m,m)^{(1)}$}
\label{Bmm}
The simple roots of $B(m,m)$ are:
$(\ve_i-\d_i),\,i=1,\ldots,m;\,(\d_i-e_{i+1}),\,i=1,\ldots,m-1;\,\d_m$.
As matrix realization of the simple roots we take: 
\bea 
E_{\ve_i-\d_i} &=& e_{4m{+}2{-}i,2m{+}2{-}i}+e_{i,2m{+}1{+}i} \nn
E_{\d_i-\ve_{i{+}1}} &=& e_{2m{+}1{-}i,4m{+}2{-}i}-e_{2m{+}1{+}i,i{+}1}
\nn 
E_{\d_m} &=& e_{m{+}1,3m{+}2}-e_{3m{+}1,m{+}1} 
\ena
while the matrix realization of the affine root is 
$$
\l e_{-\th} = \l(e_{2m{+}1,2m{+}2} + e_{4m{+}1,1})
$$
where the highest root is $\th=(\ve_1+\d_1)$. $\Psi$ takes the form of
eq. \rf{psi} with: 
$$
\psi = \left ( \begin{array}{ccc|clcc}
1&&&&&&\\
&\ddots&&&&&\\
&&1&&&&\\ \hline
&&&1&&&\\ \hline
&&&0&1&&\\
&&&&\ddots&\!\!\!\!\ddots&\\
&&&&&&1\\
\l&&&&&&0 \end{array} \right )  \hspace{20mm} 
\tilde{\psi} = \left ( \begin{array}{clcr|c|ccc}
0&\sm 1&&&&&&\\
&\ddots&\!\!\!\!\ddots&&&&&\\
&&&\sm 1&&&&\\
&&&0&\sm 1&&&\\ \hline 
&&&&&1&&\\
&&&&&&\ddots&\\
\l&&&&&&&1 
\end{array} \right ) 
$$
where the blocks are $m\times m$-matrices. 

\subsubsection{$D(m{+}1,m)^{(1)}$}
\label{Dx}

As simple roots of $D(m{+}1,m)$ we take: 
$(\ve_i-\d_i),\,i=1,\ldots,m;\,
(\d_i-\ve_{i{+}1}),\,i=1,\ldots,m;\,(\d_m+\ve_{m{+}1})$. The matrix
realization is:
\bea 
E_{\ve_i-\d_i}& = & e_{i,2m{+}2{+}i} + e_{4m{+}3{-}i,2m{+}3{-}i}\nn
E_{\d_i-\ve_{i{+}1}}&=& 
e_{2m{+}2{-}i,4m{+}3{-}i}- e_{2m{+}2{+}i,i{+}1}\nn 
E_{\d_m+\ve_{m{+}1}}&=& e_{m{+}1,3m{+}3}- e_{3m{+}2,m{+}2}
\ena
While the realization of the affine root is 
$E_{-\th}=\l (e_{2m{+}2,2m{+}3}+e_{4m{+}2,1}$, where the highest root
is $\th=\ve_1+\d_1$. In this realization, $\Psi$ takes the form of
eq. \rf{psi} with: 
$$
\psi = \left ( \begin{array}{ccc|clcc}
1&&&&&&\\
&\ddots&&&&&\\
&&1&&&&\\ \hline \vspace{-1.5mm} 
&&&1&&&\\ 
&&&1&&&\\ \hline
&&&0&1&&\\
&&&&\ddots&\!\!\!\!\ddots&\\
&&&&&&1\\
\l&&&&&&0 \end{array} \right ) \hspace{20mm}
\tilde{\psi} = \left ( \begin{array}{clcr|cc|ccc}
0&\sm 1&&&&&&&\\
&\ddots&\!\!\!\!\ddots&&&&&&\\
&&&\sm 1&&&&&\\
&&&0&\!\sm\! 1\!\!&\!\!\sm\! 1\!\!&&&\\ \hline
&&&&&&1&&\\
&&&&&&&\ddots&\\
\l&&&&&&&&1\end{array}\right ) 
$$
where the `big' blocks are $m\times m$-matrices. 

\subsubsection{$D(m,m)^{(2)}$} 

In this case, the invariant algebra is not a regular subalgebra of
$D(m,m)$, which makes the simple root system slightly more
complicated. 
We give first the fermionic simple roots of $D(m,m)$: 
$(\ve_i-\d_{i+1}),\,i=1,\ldots,m{-}1;\,
(\d_i-\ve_i),\,i=1,\ldots,m;\,\d_m+\ve_m$. 
The matrix realization is: 
\bea 
E_{\ve_i-\d_{i+1}} &=& e_{4m{-}i,2m{+}1{-}i} + e_{i,2m{+}1{+}i} \nn
E_{\d_i-\ve_i} &=& e_{2m{+}1{-}i,4m{+}1{-}i} - e_{2m{+}i,i} \nn
E_{\d_m+\ve_m} &=& e_{m,3m{+}1} - e_{3m,m{+}1} 
\ena 

The automorphism $\t$ is the diagram-automorphism 
$\ve_m \rightarrow -\ve_m$, and a simple root system of the
invariant subalgebra is found by replacing 
$E_{\d_m+\ve_m}$ (which is not invariant under $\t$) by 
$$
E_{\d_m+\ve_m}+E_{\d_m-\ve_m}=
e_{m,3m{+}1} +e_{m{+}1,3m{+}1} - e_{3m,m{+}1} - e_{3m,m{+}2}
$$
The realization of the `affine root' is 
$\l(e_{m,2m{+}1}-e_{m{+}1,2m{+}1}-e_{4m,m}+e_{4m,m{+}1})$. $\Psi$
takes the form of eq. \rf{psi} with: 
$$
\psi = \left ( \begin{array}{rlcc|ccc}
0&1&&&&&\\
&\ddots&\!\!\!\!\ddots&&&&\\
&&&1&&&\\
\l&&&0&1&&\\ \hline 
\sm\l&&&&1&&\\
&&&&&\ddots&\\
&&&&&&1   \end{array} \right )   \hspace{20mm} 
\tilde{\psi} = \left ( \begin{array}{ccc|rlcc}
\sm 1&&&&&&\\
&\ddots&&&&&\\
&&\sm 1&\sm 1&&&\\ \hline 
&&&0&1&&\\
&&&&\ddots&\!\!\!\!\ddots&\\
&&&&&&1\\
&&\sm\l&\l&&&0   \end{array} \right ) 
$$
where the blocks are $m\times m$-matrices. 

\subsubsection{$A(2m,2m)^{(4)}$}

It is convenient to split $A(2m,2m)$ into an $A(2m,2m{-}1)$
subalgebra plus an extra row an column. The invariant subalgebra is
then a $B(m,m)=osp(2m{+}1,2m)$ subalgebra of $A(2m,2m{-}1)$. 

We have already found the simple roots of $B(m,m)$ in subsection
\ref{Bmm}. We use the same basis here, supplemented with the affine
root $\l (e_{2m{+}1,4m{+}2}+e_{4m{+}2,1})$.  
In this realization, $\Psi$ takes the form of
eq. \rf{psi} with: 
$$
\psi = \left ( \begin{array}{ccc|clcc|c}
1&&&&&&&\\
&\ddots&&&&&&\\
&&1&&&&&\\ \hline 
&&&1&&&&\\ \hline 
&&&0&1&&&\\
&&&&\ddots&\!\!\!\!\ddots&&\\
&&&&&&1&\\
&&&&&&0&\l \end{array} \right ) \hspace{20mm}
\tilde{\psi} = \left ( \begin{array}{clcr|c|ccc}
0&\sm 1&&&&&&\\
&\ddots&\!\!\!\!\ddots&&&&&\\
&&&\sm 1&&&&\\
&&&0&\sm 1&&&\\ \hline 
&&&&&1&&\\
&&&&&&\ddots&\\
&&&&&&&1\\ \hline 
\l&&&&&&& \end{array}\right ) 
$$
where the blocks are $m\times m$-matrices. 

\sect{Ker(ad $\L$)} 



\subsection{Construction of elements in $\ck_\bo$}
\label{App}

In this subsection we describe the construction of the elements of
$\ck_\bo$. 
We will use a matrix realization of $\ca$ of the form 
$$
x=x_\bz+x_\bo\in\ca=\ca_\bz\oplus\ca_\bo: \;\;\;\; 
x = \left ( \begin{array}{cc}a & \a \\ \b & b\end{array} \right ) \;\;
x_\bz = \left ( \begin{array}{cc}a & 0 \\ 0 & b\end{array} \right ) \;\;
x_\bo = \left ( \begin{array}{cc}0 & \a \\ \b & 0\end{array} \right ) 
$$

\begin{table}
\begin{center}
\begin{tabular}{|r|l|}
\hline 
Affine algebra & Exponents \\ \hline 
$A_{m}^{(1)}$ & $1,2,3,\ldots,m $ mod $m+1$\\  
$A_{2m}^{(2)}$ & $1,3,5,\ldots,2m-1,2m+3,\ldots, 4m+1$ mod $4m+2$ \\ 
$A_{2m{-}1}^{(2)}$ & $1,3,5,\ldots, 4m-3$ mod $4m-2$\\
$B_{m}^{(1)}$,  $C_{m}^{(1)}$ & $1,3,5,\ldots, 2m-1$ mod $2m$\\
$D_{m}^{(1)}$ & $1,3,5,\ldots, 2m-3,m-1$ mod $2m-2$ \\
$D_{m+1}^{(2)}$ & $1,3,5,\ldots, 2m+1 $ mod $2m+2$ \\
\hline 
\end{tabular}
\end{center}
\caption{Exponents of the affine algebras that provide the bosonic
part of the affine Lie superalgebras with a fermionic simple root
system.}
\label{tbexp}
\end{table}

We write 
$$
\Psi = \left ( \begin{array}{cc} 0 & \psi \\ \tilde{\psi} & 0
\end{array} \right ) 
$$
and we find 
$$
\L =  [\Psi,\Psi] = 
\left ( \begin{array}{cc} \psi\tilde{\psi} & 0 \\ 0 &
\tilde{\psi}\psi \end{array} \right ) \;\; = \;\;
\left ( \begin{array}{cc} \L_1 & 0 \\ 0 & \L_2 \end{array} \right ) 
$$
where $\L_i$ are principal elements in $\cg_i$. $\ck_\bz$ consists of
elements 
$$
X = \left ( \begin{array}{cc} x_1 & 0 \\ 0 & x_2 \end{array} \right ) 
$$ 
where $x_i\in\ker(\ad(\L_i))$, or, in the case of
$A(2m{-}2,2m{-}1)^{(2)}$, of elements $X+u$, where $u$ is an element
of the $U(1)^{(1)}$ subalgebra of $A(2m{-}2,2m{-}1)^{(2)}$. 

$\ker(\ad(\L_i))$ are well known, and are generated by the elements 
$\L_i^{e_j}$, where $\{e_j\}$ is the set of exponents of $\cg_i$, with
the following exceptions: 
\begin{itemize}
\item[1)] $D_m^{(1)}$ has exponents 
$n=1,3,5,\ldots,2m-3$ and $m-1$ mod $2m-2$. 
There is an element $\tilde{c}_{2m-2}$ of $\ck_\bz$ corresponding to 
the spurious exponent $m-1$, which is not associated to $\L^{m-1}$. 
\item[2)] $A_{2m{-}1}^{(2)}$ has exponents $n=1,3,5,\ldots,4m{-}3$
mod $4m{-}2$. The element of $\ck_\bz$ corresponding to the exponent 
$2m{-}1$ is $\l^{2m{-}1}-{1\over 2m} \tr(\l^{2m{-}1}) \one$. 
\end{itemize}
We have collected the exponents of the relevant bosonic affine Lie
algebras in table~\ref{tbexp}, and we refer to
\cite{DrSo,KBOOK} for further details.~\label{exponents} 

Taking commutators $[\Psi,\ck_\bz]$,
$[[\Psi,\ck_\bz],\ck_\bz]$, etc.~\footnote{In the particular case of
$A(1,1)^{(1)}$, one has to add the odd generator
$\bar{\Psi} = \left ( \begin{array}{cc} 0 & \psi \\ -\tilde{\psi} & 0
\end{array} \right )$
of principal grade~1, together with $[\bar{\Psi},\ck_\bz]$,
$[[\bar{\Psi},\ck_\bz],\ck_\bz]$, etc.} we can now construct
more elements of
$\ck_\bo$. The elements found in this way were summarized on page
\pageref{ferm-elts}. In table \ref{tb3} we compare the number of these
generators, modulo twice the Coxeter number, with the maximal possible
number 2 min($d_+,d_-$) found in the previous subsection.  
\begin{table}
\begin{center}
\begin{tabular}{|r|r|r|}
\hline 
Superalgebra & & 2 min($d_+,d_-$) \\ \hline 
$A(m,m)^{(1)}$ & $2m{+}2$ & $2m{+}2$ \\  
$A(2m,2m)^{(4)}$ & $4m{+}2$ & $4m{+}2$ \\ 
$A(2m{-}1,2m{-}1)^{(2)}$ & $4m{-}2$ & $4m$ \\
$A(2m{-}2,2m{-}1)^{(2)}$ & $4m{-}2$ & $4m{-}2$ \\ 
$B(m,m)^{(1)}$ & $2m$ & $2m$ \\
$D(m{+}1,m)^{(1)}$ & $2m$ & $2m$ \\
$D(m,m)^{(2)}$ & $2m$ &  $2m$ 
\\
\hline 
\end{tabular}
\end{center}
\caption{The second column gives the number of generators in
$\phi(\ck_\bo)$ of the form $\phi(\Psi)$, $\phi(\bar{\Psi})$,
$\phi([\Psi,\ck_\bz]), \phi([[\Psi,\ck_\bz],\ck_\bz])$,
$\phi([\bar{\Psi},\ck_\bz]),
\phi([[\bar{\Psi},\ck_\bz],\ck_\bz])$, etc., while the third column gives
the upper limit 2 min($d_+,d_-$) on the number of elements of
$\phi(\ck_\bo)$. As we see, the second and third columns agree in all
cases except $A(2m{-}1,2m{-}1)^{(2)}$.}
\label{tb3}
\end{table}
We see that the number of fermionic generators saturate the limit 
in all cases except $A(2m{-}1,2m{-}1)^{(2)}$. In this last case we
find by explicit calculations two extra series of fermionic generators
with grades $2m{-}1$ and $6m{-}3$, both modulo $8m{-}4$.  
\label{Akernel}
In the matrix realization given in section \ref{Ax} these extra
elements are: 
\be
\xi_{i},\> \bar{\xi}_i = \left ( \begin{array}{cc} 0 & \a_i \\ \b_i & 0
\end{array}
\right ) 
\lab{xi}
\ee
and 
\be
\a_{2m{-}1} = \left ( \begin{array}{rcccr}
0 & 0 & \cdots & 0 & 0 \\
\vdots & \vdots & & \vdots & \vdots \\
0 & 0 & \cdots & 0 & 0 \\
\l & 0 & \cdots & 0 & \sm 1 \\
\sm\l & 0 & \cdots & 0 & 1 \\
0 & 0 & \cdots & 0 & 0 \\
\vdots & \vdots & & \vdots & \vdots \\
0 & 0 & \cdots & 0 & 0 \end{array} \right ) \hspace{20mm}
\b_{2m{-}1} = \left ( \begin{array}{rrrrrrrr}
0 & \cdots & 0 & \sm 1 & 1 & 0 & \cdots & 0 \\
0 & \cdots & 0 & 0 & 0 & 0 & \cdots & 0 \\
\vdots & & \vdots & \vdots & \vdots & \vdots & & \vdots \\
0 & \cdots & 0 & 0 & 0 & 0 & \cdots & 0 \\
0 & \cdots & 0 & \l & \sm\l & 0 & \cdots & 0 \end{array} \right ) 
\label{ab}
\ee
and in addition: 
\bea
\a_{2m{-}1+n(8m{-}4)} & = & \l^{2n} \a_{2m{-}1} \nn
\b_{2m{-}1+n(8m{-}4)} & = & \l^{2n} \b_{2m{-}1} \nn
\a_{6m{-}3+n(8m{-}4)} & = & \l^{2n+1} \a_{2m{-}1} \nn
\b_{6m{-}3+n(8m{-}4)} & = & -\l^{2n+1} \b_{2m{-}1} \nn
\ena 

\subsection{Commutation Relations} 
\label{sect:comrel}

$\ck_\bz$ is an Abelian algebra, so we need to find only 
the commutation relations $[\ck_\bo,\ck_\bo]$ and $[\ck_\bz,\ck_\bo]$. 

Remarkably, we find that we can always choose the elements
$\psi_{-1+2i}$ and $\bar{\psi}_{1+2i}$ to be of the following simple form:  
\bea
\psi_{-1+2i} & = & \left ( \begin{array}{cc} 
0 & \psi(\tilde{\psi}\psi)^{i-1} \\
\tilde{\psi}(\psi\tilde{\psi})^{i-1} & 0 \end{array} \right )  \nn
\bar{\psi}_{1+2i} & = & \left ( \begin{array}{cc} 
0 & \psi(\tilde{\psi}\psi)^{i} \\
-\tilde{\psi}(\psi\tilde{\psi})^{i} & 0 \end{array} \right )
\ena 
Note that $\psi(\tilde{\psi}\psi)^i = \psi\L_2^i = \L_1^i\psi$ and
$\tilde{\psi}(\psi\tilde{\psi})^i = \tilde{\psi}\L_1^i = 
\L_2^i\tilde{\psi}$. 
One can easily verify that the commutation relations of these
generators are the ones given in equation~\rf{comrel}. 
In the case of the algebra $A(2m{-}1,2m{-}1)^{(2)}$, where we have
extra fermionic generators $\xi_i$ and $\bar{\xi}_i$, explicit
calculations give the  commutation relations eq. \rf{comrel2}. 

In general $\ck^{2i}$ is generated by $c_{2i} = \L_1^i+0$~\footnote{In
order to simplify the notation, and when the meaning is otherwise
clear, we will sometimes write elements 
$\left ( \begin{array}{cc} a & 0 \\ 0 & b \end{array}\right )$ in the
form $a+b$.}
and $d_{2j} = 0 + \L_2^j$ 
where $i$ and $j$ are exponents of respectively $\cg_1$ and $\cg_2$,
and $\cz^{2i}$ by $b_{2i} = c_{2i} + d_{2i}$. 
The exceptions, which follows directly from the exceptions in the
bosonic case described in the previous section, are:
\begin{itemize} 
\item[1)] For the algebra $D(m{+}1,m)^{(1)}$ there is an `extra' 
element $\tilde{c}_{2m}+0\in\ck_\bz$ which is not related to powers of $\L_i$. 
In the matrix realization given in section \ref{Dx}, $\tilde{c}_{2m}$ has the
same form as $\b_{2m{-}1}$ in equation (\ref{ab}). 
\item[2)] For the algebra $A(2m{-}1,2m)^{(2)}$, $\ck^{4m{-}2}$ is
generated by 
$$
c'_{4m{-}2} 
= \left (\L_1^{2m{-}1}-{1\over 2m} \tr(\L_1^{2m{-}1})\one \right )
+ 0
$$ 
and the
$u(1)$ generator 
$$
d'_{4m{-}2}={1\over 2m}\one + {1\over 2m{-}1}\one$$ 
\item[3)] For the algebra $A(2m{-}1,2m{-}1)^{(2)}$, $\ck^{4m{-}2}$ is
generated by 
$$
c'_{4m{-}2}= \left ( 
\L_1^{2m{-}1}-{1\over 2m} \tr(\L_1^{2m{-}1})\one \right ) + 0
$$ 
and 
$$
d'_{4m{-}2}=0+\left (
\L_2^{2m{-}1}-{1\over 2m} \tr(\L_2^{2m{-}1}) \right ) 
$$ 
\end{itemize} 

Using these expressions, it is straightforward to
find which elements of $\ck_\bz$ are in the centre. 
The only exception is the `extra' element in $\ck_\bz$ 
in the case of $D(m{+}1,m)^{(1)}$. Here, a simple calculation shows
that $\tilde{c}_{2m}\psi = \tilde{\psi} \tilde{c}_{2m} = 0$, which shows that
$\tilde{c}_{2m}+0\in\cz$. 
\label{Dcentre}

It is easy to verify by direct calculation that 
\bea
{}[c_{2i},\psi_{-1+2j}] &=& \bar{\psi}_{1+2(i{+}j{-}1)} \nn
{}[c_{2i},\bar{\psi}_{1+2j}] &=& -\psi_{-1+2(i{+}j{+}1)} 
\ena 
while the commutation relations of $d_{2i}$ follow directly by
observing that $c_{2i}+d_{2i}\in\cz$. 
Using an explicit realization of $\xi_i$ and $\bar{\xi}_i$ we find also 
\be 
{}[c_{2i},\xi_{j}] = [c_{2i},\bar{\xi}_{j}]= 0 \nn
\ee
Moreover, a direct calculation shows that in the algebras
$A(2m{-}1,2m)^{(2)}$ and $A(2m{-}1,2m{-}1)^{(2)}$ it is possible to
choose the normalizations
$c_{4m{-}2}=a_{4m{-}2}\, c'_{4m{-}2}$ and 
$d_{4m{-}2}=b_{4m{-}2}\, d'_{4m{-}2}$ such that these elements also
satisfy the commutation relations given above. 


\end{document}